\documentclass{LMCS}

\def\dOi{11(2:3)2015}
\lmcsheading%
{\dOi}
{1--24}
{}
{}
{Dec.~29, 2013}
{May.~31, 2015}
{}

\ACMCCS{[{\bf Theory of computation}]: Theory and algorithms for
  application domains---Machine learning theory; Logic---Modal and
  temporal logics; [{\bf Software and its engineering}]: Software organization and properties---Software functional properties---Formal methods---Software verification}


\usepackage{amssymb}
\usepackage{amsmath}
\usepackage{graphicx}
\usepackage{color}
\usepackage{subfigure}
\newcommand{\guido}{}
\newcommand{\luca}{}

\newcommand{\bbR}{\mathbb{R}}
\newcommand{\bbE}{\mathbb{E}}

\newcommand{\true}{\mathtt{tt}}
\newcommand{\false}{\mathtt{ff}}
\newcommand{\until}[2]{\mathbf{U}_{[#1,#2]}}
\newcommand{\eventually}[2]{\mathbf{F}_{[#1,#2]}}
\newcommand{\always}[2]{\mathbf{G}_{[#1,#2]}}

\renewcommand{\phi}{\varphi}

\theoremstyle{plain}

\renewcommand{\vec}[1]{\mathbf{#1}}
\newcommand{\bs}[1]{\boldsymbol{#1}}

\begin{document}

%\mainmatter

\title{Learning and designing stochastic processes from logical constraints}

\author[L.~Bortolussi]{Luca Bortolussi\rsuper a}	%required
\address{{\lsuper a}Modelling and Simulation Group, Saarland University, Germany \and Department of Mathematics and  Geosciences, University of Trieste \and 
CNR/ISTI, Pisa, Italy}	%required
%\email{luca@dmi.units.it}  %optional
\thanks{{\lsuper a}Work partially supported by EU-FET project QUANTICOL (nr. 600708) and by FRA-UniTS.}	

\author[G.~Sanguinetti]{Guido Sanguinetti\rsuper b}	%optional
\address{{\lsuper b}School of Informatics, University of Edinburgh \and
 SynthSys, Centre for Synthetic and Systems Biology, University of Edinburgh}	%optional
%\email{}  %optional
\thanks{{\lsuper b}Work supported by European Research Council under grant MLCS 306999.}	%optional

\begin{abstract}
Stochastic processes offer a flexible mathematical formalism to model and reason about systems. Most analysis tools, however, start from the premises that models are fully specified, so that any parameters controlling the system's dynamics must be known exactly. As this is seldom the case, many methods have been devised over the last decade to infer (learn) such parameters from observations of the state of the system. In this paper, we depart from this approach by assuming that our observations are {\it qualitative} properties encoded as satisfaction of linear temporal logic formulae, as opposed to quantitative observations of the state of the system. An important feature of this approach is that it unifies naturally the system identification and the system design problems, where the properties, instead of observations, represent requirements to be satisfied. We develop a principled statistical estimation procedure based on maximising the likelihood of the system's parameters, using recent ideas from statistical machine learning. We demonstrate the efficacy and broad applicability of our method on a range of simple but non-trivial examples, including rumour spreading in social networks and hybrid models of gene regulation.  
\end{abstract}

\maketitle

%%%%%%%%%%%%%%%%%%%%%%%%%%%%%%%%%%%%%%%%%%%%%%%%
%%%%%%%%%%%%%%%%%%%%%%%%%%%%%%%%%%%%%%%%%%%%%%%%
%%%%%%%%%%%%%%%%%%%%%%%%%%%%%%%%%%%%%%%%%%%%%%%%

\section{Introduction}

Stochastic processes are fundamental tools for modelling and reasoning about many physical and engineered systems. Their elegant mathematical formulation allows to capture quantitatively the mechanisms underlying the intrinsically noisy dynamics frequently encountered in many applications, ranging from computer networks to systems biology. At the same time, their importance has motivated intense research in analytical and computational tools to characterize emergent properties of models, and to \guido{efficiently simulate system trajectories by sampling from stochastic processes}. \guido{While the predictive power of stochastic models is a key to their success in scientific applications, the development of algorithms and methodologies to reason about stochastic models has been a consistent focus of research in theoretical computer science over the past five decades.} Of particular importance in verification, \guido{and for this paper,} is (stochastic) model checking: given a property (formalised as a formula in a suitable logic), estimate the probability that it is satisfied by a random trajectory of the model \cite{Baier2008}. 

\luca{Model checking tools, either numerical or statistical}, however, can only be deployed if a model is fully specified (or, at least, if sample trajectories can be computed effectively). This requirement is often conceptually and practically untenable in many situations: modelling is the result of a mathematical formalisation of scientific expertise, and \guido{while such expertise is often able to define suitable model structures, it is implausible to expect to be able} to pin-point uniquely defined values for the many parameters which are involved  in many complex models. The increasing awareness of this limitation has motivated considerable research in statistical machine learning and systems engineering; while parameter synthesis is still an open research question, there are several approaches which estimate the parameters of a stochastic process from observations of \luca{the state of the modelled system.} These approaches assume that (noisy) observations of the actual state of the system are available, usually in the form of time series \cite{Andreychenko:approximate12,Opper:variational07}. 

In this paper, we shift the focus from observations of the state of the system to observations of the {\it emergent properties} of the system:  \luca{we assume to observe {\it truth values} or {\it satisfaction probabilities} of logical formulae over sample trajectories of the system, and use such data to identify the parameters of the stochastic process.}
%the data we wish to use \luca{to identify the parameters} of the stochastic process are directly {\it truth values} or {\it satisfaction probabilities} of logical formulae over sample trajectories of the system. 
The rationale for exploring this problem, that to our knowledge has not been extensively studied (see below for related work), is three-fold: in the first instance, in many applications gathering and storing \guido{(multiple}) time series data is difficult and expensive, while qualitative global properties (e.g. phenotypes in a biological application) may be more readily available. Secondly, learning a model from logical constraints more closely matches the modelling process itself: generally a suitable model is chosen to capture some qualitative behaviour of the system (e.g. a negative feedback loop for an oscillator); it is therefore natural to also attempt to recover plausible parametrisations from such data. Thirdly, this approach illustrates the close relationship between the system identification and the system design problem: one could equally well imagine the satisfaction probabilities to be not the result of observations, but requirements set out by a user which need to be matched.

Solving these problems presents considerable computational and statistical challenges: in order to define a suitable objective function for parameter optimisation (e.g. a likelihood function), one needs to be able to explicitly determine the functional dependence of satisfaction probabilities on the parameters, which is impossible in all but the simplest cases. One can however obtain an approximate estimate of this likelihood at specific parameter values by using a Statistical Model Checking (SMC) procedure. This enables us to leverage a powerful class of machine learning algorithms for optimising unknown (but computable) functions, Bayesian Optimisation. Within the Bayesian Optimisation family, we select a provably convergent, recently developed global optimisation algorithm, the Gaussian Process Upper Confidence Bound (GP-UCB) optimisation algorithm. We show that this approach is effective and accurate in both the system identification and the system design problems on a range of non-trivial examples.

The rest of the paper is organised as follows: we start by briefly recapitulating the fundamental notions about stochastic processes and temporal logics. We then introduce the main methodological tools underpinning our approach. The approach is evaluated on a number of model examples, including continuous-time Markov chains and hybrid stochastic systems. We conclude by discussing the merits and limitations of this approach.

\subsection*{Related work} 
This paper grows out of a conference paper of the same title \cite{Bortolussi2013}. While the core idea is the same, it is extended in several directions: we now apply the methodology to a broader class of stochastic processes (including SDEs and hybrid models), we improve the algorithm by incorporating a hyper-parameter optimisation routine, we provide approximate estimates of the uncertainty over the optimal parameters, and we devise methodology to handle the non-homogeneous nature of the noise in SMC. Furthermore, the paper is completed by a  new experimental section on different examples. Within the recent literature, earlier attempts were made to use model checking methods for parameter estimation in \cite{Donaldson2008}; while the underlying idea of constraining a model with logical properties is shared, the quantitative semantics we employ here and the more powerful algorithmic solutions lead to considerable differences. Also related is the idea of model repair \cite{Bartocci2011}, whereby the parametrisation of an original model is locally modified to increase the satisfaction probability of a logical property. However, this approach is based on parametric model checking \cite{Hahn2011}, which heavily suffers from state space explosion.

Optimisation methods can be fruitfully employed in other formal modelling scenarios: \cite{Bartocci2013} uses similar algorithmic procedures to optimise the robustness with which a formula is satisfied, while \cite{Bartocci:data14} attacks the converse problem of identifying properties with high satisfaction probability within a parametric family of formulae (given a fixed model). 

Within the machine learning literature, \cite{Cseke:approximate13} has developed novel approximation techniques to solve the problem of {\it Bayesian inference} from (continuous-time) constraints on trajectories. The considerably harder nature of this problem (involving estimation of a whole posterior process, as opposed to just the parameters) implied however that only a very restricted class of models and constraints could be considered in that paper.

%%%%%%%%%%%%%%%%%%%%%%%%%%%%%%%%%%%%%%%%%%%%%%%%
%%%%%%%%%%%%%%%%%%%%%%%%%%%%%%%%%%%%%%%%%%%%%%%%
%%%%%%%%%%%%%%%%%%%%%%%%%%%%%%%%%%%%%%%%%%%%%%%%

%\red{RICORDARSI DI CITARE MODEL REPAIR DI EZIO E HSB NOSTRO}

\section{Background}
\label{sec:background}
\guido{ In this section, we provide a brief introduction to the fundamental mathematical and logical concepts underpinning our approach. We will start in Section \ref{sec:sp} by briefly recalling the broad class of systems we consider.
 We then introduce in Section \ref{sec:MITL} the logical formalism in which system properties will be encoded, namely Metric Interval Temporal Logic (MiTL). We stress that this particular choice of logic is not essential for our approach, which will work for any logic whose predicates are verified on individual, time bounded trajectories. Once these preliminaries are established, we will formally define the system identification problem (Section \ref{sec:sysIdent}) and the system design problem (Section \ref{sec:sysDesign}). In both cases, we limit ourselves to the problem of identifying parameters of a model with fixed structure, leaving structural identification to further work.}

%
%\red{Define the problem, by introducing the notion of stochastic process of interest in more detail, and by describing the identification and the design problem when we have access to the TRUE probability distributions. Anticipate at this level the definition of the logic, and of the likelihood function/ JS divergence}

\subsection{Stochastic Processes}
\label{sec:sp}
Here we provide a quick and informal introduction to the classes of stochastic processes considered in this paper, briefly introducing the simulation algorithms used for drawing samples from them. The reader interested in a more thorough introduction is referred to standard textbooks like \cite{Gardiner:handbook02,Oksendal2003,bujorianu_analysis_2012}.

Let the state of the system at any one time $t\in[0,T]$ be defined by a {\it state variable} $\mathbf{V}$ taking values in a suitable measurable space $\mathcal{D}$. A {\it stochastic process} is a family of $\mathcal{D}$-valued random variables indexed by $t$; equivalently, this defines a measure over the space of trajectories of the system $\mathcal{T}=\{f\colon[0,T]\rightarrow\mathcal{D}\}$. Selecting a finite subset of indices $t_0,\ldots,t_N$, one obtains finite-dimensional random variables given by the configurations of the system at those times; the distribution of such random variables are the {\it finite-dimensional marginals} of the process. The process is {\it Markovian} if, given any finite set of state values $\mathbf{V}(t_1),\ldots,\mathbf{V}(t_N)$, the finite-dimensional joint marginal distribution factorises as\begin{equation}
p\left(\mathbf{V}(t_1),\ldots,\mathbf{V}(t_N)\right)=p\left(\mathbf{V}(t_1)\right)\prod_{j=2}^Np\left(\mathbf{V}(t_j)\vert\mathbf{V}(t_{j-1})\right).
\label{MarkovProp}\end{equation}
The conditional probability $p\left(\mathbf{V}(t+\delta t)\vert\mathbf{V}(t)\right)$ is usually termed transition probability; its derivative (transition rate) is called the generator of the Markov process. We will assume that the parametric dependence of the system is contained in the generator of the Markov process, and that the generator does not explicitly depend on time (time homogeneous process). %\red{STIAMO ESCLUDENDO LA COSTANTE DI DIFFUSIONE QUI?} 
The Markov property implies that the transition probabilities satisfy  deterministic differential equations which in general are known as {\it Chapman-Kolmogorov equations}. We will consider the following three types of Markovian stochastic processes:\begin{itemize}
\item {\it Continuous-Time Markov Chains (CTMCs)}
CTMCs are a common mathematical model of stochastic dynamical processes in many areas of science; they are Markovian stochastic processes with discrete state space (i.e. $\mathcal{D}\subset\mathbb{Z}^d$).
%, including computer science, where they are heavily used in performance analysis, and computational biochemistry and biology, where they are the  standard dynamical model of biochemical reactions. 
We will adopt the  population view of CTMCs \cite{bortolussi_continuous_2013}, in which the state space is described  by a collection of $n$ integer-valued variables $\vec{V} = (V_1,\ldots,V_n)$, describing the number of entities in each population of the model, and will borrow the notation of chemical reactions 
\cite{van_kampen_stochastic_2007} to describe CTMCs. The transition probability of a CTMC obeys the {\it Chemical Master Equation} (CME), a (potentially infinite) set of coupled ordinary differential equations. The CME cannot be solved in all but the simplest cases; however, an exact algorithm, Gillespie's Stochastic Simulation Algorithm, exists to draw samples from a (time homogeneous) CTMC \cite{SB:Gillespie:1977:gillespieAlgorithm}.

\item{\it Stochastic Differential Equations (SDEs)}
SDEs \cite{Oksendal2003} define stochastic processes with continuous state space (usually $\mathcal{D}=\mathbb{R}^n$) and continuous (but nowhere differentiable) trajectories. We can think of SDEs as ordinary differential equations associated with a vector field which is randomly perturbed at each point by a white noise process. SDEs play an important role in science and engineering; in recent years, they have attracted considerable attention in computer science as fluid approximations to CTMCs. We will confine ourselves to It\^o SDEs, which can be written as 
\[ d\vec{V} = F(\vec{V})dt + G(\vec{V}) d\vec{W}, \]
where $\vec{W}$ is a $d$-dimensional Wiener process (whose derivative is known as white noise), $F$ is the $n$-dimensional \emph{drift} function and $G$ is the $n\times d$ diffusion matrix. 
SDEs can be simulated using a variety of numerical schemes; here we will use the  Euler-Maruyama scheme, which fixes a time step $h$ and iteratively computes $\vec{v}(t+h) = \vec{v}(t) +  F(\vec{v}(t))h +  G(\vec{v}(t)) \mathcal{N}(0,h I_d)$, where $\mathcal{N}(\vec{0},h I_d)$ is a $d$-dimensional Gaussian random variable with mean $\vec{0}$ and diagonal covariance matrix $h I_d$. It is important to notice that, in contrast to the CTMC case, this simulation procedure is no longer exact, but introduces an error which reduces to zero only in the $h\rightarrow 0$ limit.

% If we have a description of a model in terms of reactions, we can costruct an SDE as follows. The drift function is obtained by summing the rate functions multiplied by the update vector $\bs{\nu}$: $F(\vec{v},\theta) = \sum_{j=1}^m \bs{\nu_j}a_j(\vec{v},\theta)$, while the diffusion can be either constructed according to the recipe of the  Langevin equation~\cite{van_kampen_stochastic_2007}, or by assuming a specific form of the diffusion. In this paper, we will go along this second direction, imposing a constant diffusion, possibly different for each variable, i.e. $G(\vec{V}) = diag(\bs{\sigma})$. $\bs{\sigma} = (\sigma_1,\ldots,\sigma_n)$. Obviously, $\bs{\sigma}$ are new parameters of the system, and will be included in $\theta$.  To denote the solution of SDEs, we follow the same conventions as for CTMCs.

%\red{Controllare che non abbia scritto cazzate madornali}.
\item{\it Stochastic Hybrid Systems (SHS)}
More generally, one may also consider stochastic processes with hybrid state space, i.e. $\mathcal{D}\subset\mathbb{Z}^d\times\mathbb{R}^D$. These may arise e.g. as approximations to CTMCs where some of the populations have large numbers (which can be well approximated as continuous variables) while others have sufficiently small numbers to require a discrete treatment \cite{Bortolussi2010,Bortolussi2013b,Ocone2013}. The models so obtained are known as stochastic hybrid systems (SHS)~\cite{bujorianu_analysis_2012}, and their dynamics can be seen as a sequence of discrete jumps, instantaneously modifying population variables, interleaved by periods of continuous evolution along a trajectory of the SDE. As the rates of discrete transitions can depend on the continuously evolving variables, and vice versa, these systems can exhibit rich dynamics; nonetheless, the Markovian nature of the SHS still implies that effective (approximate) simulation algorithms can be obtained, see for instance \cite{riley_simulation_2008}.
\end{itemize}
\subsection{Metric interval Temporal Logic}
\label{sec:MITL}

We will consider properties of stochastic trajectories specified by Metric interval Temporal Logic (MiTL),  see \cite{MC:AlurHenzinger:ACM1996:MiTL,MC:Maler:FORMATS2004:MiTL}. 
This logic is a linear temporal logic, so that the  truth of a formula can be assessed over single trajectories of the system. MiTL, in particular, is used to reason on real-time systems, like those specified by CTMC or SDEs. Here we consider the fragment of MiTL in which all  temporal operators are all time-bounded; this choice is natural in our context, as we want to compare a model with properties of experimental observations of \emph{single time-bounded realisations} (essentially, time-bounded samples from its trajectory space). %Hence, MiTL is the natural choice to formalise the qualitative outcome of experiments.

The syntax of MiTL is given by the following grammar:
\[ \phi ::= \mathtt{tt}~|~\mu~|~\neg\phi~|~\phi_1\wedge\phi_2~|~\phi_1\until{T_1}{T_2}\phi_2,   \]
where $\true$ is the true formula, conjunction and negation are the standard boolean connectives, and there is only one temporal modality, the time-bounded until $\until{T_1}{T_2}$, where $T_1< T_2$ are the time bounds.  Atomic propositions  $\mu$ are defined like in Signal Temporal Logic (STL \cite{MC:Maler:FORMATS2004:MiTL}) as boolean predicate transformers: they take a real valued function $\vec{v}(t)$, $\vec{v}:[0,T]\rightarrow\bbR^n$, as input, and produce a boolean signal $s(t) = \mu(\vec{v}(t))$ as output, where $s:[0,T]\rightarrow \{\true,\false\}$. As customary, boolean predicates $\mu$ are (non-linear) inequalities on vectors of $n$ variables, that are extended point-wise to the time domain. 
Temporal modalities like time-bounded eventually and always can be derived in the usual way from the until operator: $\eventually{T_1}{T_2}\phi \equiv \true\until{T_1}{T_2} \phi$ and $\always{T_1}{T_2}\phi\equiv \neg \eventually{T_1}{T_2}\neg\phi$. 

A MiTL formula is interpreted over a real valued function of time $\vec{v}$, and its satisfaction relation is  given in a standard way, see e.g. \cite{MC:AlurHenzinger:ACM1996:MiTL,MC:Maler:FORMATS2004:MiTL}. We report here  the semantic rules for completeness:
\begin{itemize}
\item $\vec{v},t \models \mu$ if and only if $\mu(\vec{v}(t)) = \true$;
\item $\vec{v},t \models \neg\phi$ if and only if $\vec{v},t \not\models \phi$;
\item $\vec{v},t \models \phi_1\wedge\phi_2$ if and only if $\vec{v},t \models \phi_1$ and $\vec{v},t \models \phi_2$;
%\item $\vec{v},t \models \mu$ if and only if $\mu(\vec{v}(t)) = \true$;
\item $\vec{v},t \models \phi_1\until{T_1}{T_2}\phi_2$ if and only if $\exists t_1 \in [t+T_1,t+T_2]$ such that $\vec{v},t_1 \models \phi_2$ and $\forall t_0\in [t,t_1]$, $\vec{v},t_0 \models \phi_1$\footnote{For the semantics of the until, we require that at time $t_1$, both $\phi_1$ and $\phi_2$ are true, following the treatment of STL \cite{MC:Maler:FORMATS2004:MiTL}.}.
\end{itemize}\medskip

\noindent A MiTL formula $\phi$ can be verified \cite{MC:Maler:FORMATS2004:MiTL} over a real valued function $\vec v$ by first converting $\vec v$ into a vector of boolean signals $\mu_j(\vec v(t))$, where $\mu_j$ are all the atomic predicates appearing in $\phi$, and then processing these signals bottom up from the parse tree of the formula\footnote{\guido{Notice that the algorithm for monitoring a logic formula is irrelevant for the methodology presented in this paper, which only relies on the availability of boolean qualitative data.}}. 
 A standard assumption is that the so obtained boolean signals change truth value only a finite number of times in $[0,T]$ (finite variability). 
 %For CTMC that are non-explosive \cite{Gardiner:handbook02}, for instance when  rates are bounded, this holds almost surely for any $T<\infty$.

The temporal logic MiTL can be easily extended to the probabilistic setting, and interpreted over CTMC or other stochastic models like SDE or SHS~\cite{Zuliani2009,MC:Mereacre:2011:MTLmc}. Essentially, the quantity of interest is the path probability of a formula $\phi$, defined as\footnote{
We use the notation $p(x)$ to denote  the probability density of $x$ of $x$, while  $p(x\vert y)$ is used for the conditional probability density of $x$ given $y$.  
}
\[ p(\phi) = p\left(\{\vec v_{0:T}\vert \vec v_{0:T},0\models \phi\}  \right),\] i.e.\ as the probability of the set of time-bounded trajectories that satisfy the formula\footnote{We assume implicitly that $T$ is sufficiently large so that the truth of $\phi$ at time 0 can always be established from $\vec v$. The minimum of such times can be easily deduced from the formula $\phi$, see \cite{Zuliani2009,MC:Maler:FORMATS2004:MiTL}}. Here $\vec v_{0:T}$ denotes a trajectory $\vec v$ restricted to the time interval $[0,T]$.

Note that trajectories of a CTMC \guido{with bounded transition rates, or more generally  non-explosive \cite{Gardiner:handbook02},} will always enjoy the finite variability property, as they are piecewise constant and their number of jumps is finite in $[0,T]$ with probability one.  A  more complex argument, based on first passage times for Brownian motion, can be employed to show that trajectories of SDEs and SHS also have finite variability \cite{Gardiner:handbook02}.

\section{Problem definition}
\label{sec:problem}
We give here a precise definition of the two related problems we set out to solve in this work.%: given a parametric family of stochastic models and a set of logical formulae, how do we determine the model parameters that best match observed satisfaction values of the formulae on a number of independent runs (system identification)? And how can we design a stochastic process that best matches target satisfaction probabilities (system design)?

\subsection{System identification}
\label{sec:sysIdent}

Consider now a stochastic process $\vec V$ depending on a set of parameters $\theta$, and a set of $d$ MiTL formulae $\bs{\phi} = \{\phi_1,\ldots,\phi_d\}$. We assume that the truth values of the $d$ formulae have been observed over $N$ independent runs of the process and gather the observations in the $d\times N$ {\it design (or data) matrix} $D$. Given a specific value of the parameters $\theta$, the probability of observing the design matrix, $p(D\vert\theta)$, is uniquely determined, and can be computed by model checking (possibly using a randomized algorithm such as SMC). 
The \emph{system identification problem} addresses the inverse problem of finding the value(s) of parameters $\theta$ which best explain the observed design matrix. 

The key ingredient in the identification of probabilistic systems is the \emph{likelihood}, i.e. the probability of observing the data matrix $D$ for a given set of parameters $\theta$; under the assumptions that observations are independent and identically distributed the likelihood factorises as the product of the probabilities of observing the individual columns of the design matrix
\begin{equation} 
\mathcal{L}(D,\theta)=\prod_{i=1}^Np(D_i\vert\theta).\label{likelihood}
\end{equation}
The \emph{system identification problem} then corresponds to finding the parameter configuration $\theta^*$ that maximises the likelihood (\ref{likelihood}) (\emph{maximum likelihood}, ML). \luca{Equivalently, we can maximise the logarithm of $\mathcal{L}(D,\theta)$, the so called log-likelihood, as is common practice in statistics (the result is the same due to monotonicity of the logarithm).}

If prior knowledge over the parameters is available as a prior distribution $p(\theta)$ on the space of parameters $\Theta$, we can consider the un-normalised posterior distribution 
\begin{equation}
p(\theta, D)\propto p(\theta)\prod_{i=1}^Np(D_i\vert\theta).
\label{posterior}
\end{equation}
and alternatively seek to maximise this quantity, giving rise to \emph{maximum a posteriori} (MAP) estimation.

\subsection{System Design}
\label{sec:sysDesign}
Consider again $d$ MiTL  formulae  $\bs{\phi}=(\phi_1,\ldots,\phi_d)$ and a stochastic process $\vec V(t)$ depending on parameters $\theta$. We fix a \emph{target probability table} $P$ for the joint occurrence of the $d$ formulae. The \emph{system design problem} then consists of determining the parameters of the stochastic process which optimally match these probabilities.

This problem is intimately linked to system identification: in fact, one could characterise system design as {\it inference with the data one would like to have} \cite{Barnes:Bayesian11}. In our case, we are given a probability table for the joint occurrence of a number of formulae $\phi_1,\ldots,\phi_N$.\footnote{This problem formulation is different from a recent approach on parameter synthesis for CTMC using SMC, \cite{JhaTCS2011}, in which the authors look for a subset of parameters in which a single formula $\phi$ is satisfied with probability greater than $q$.} 
However, in the design case, we do not aim to use this function to estimate the likelihood of observations, rather to match (or be as near as possible to) some predefined values. We therefore need to define a different objective function that measures the distance between two probability distributions; we choose to use the Jensen-Shannon Divergence (JSD) due to its information theoretic properties and computationally good behaviour (being always finite) \cite{Cover:elements06}. This is defined as \[
JSD(p\Vert q)=\frac{1}{2}\sum_{i}\left[p_i\log\frac{2p_i}{p_i+q_i}+q_i\log\frac{2q_i}{p_i+q_i}\right]\]
where $p$ and $q$ are two probability distributions over a finite set. The Jensen-Shannon divergence is symmetric and always non negative, being zero if and only if $q=p$. Hence, system design corresponds to finding the parameter configuration $\theta^*$ that  minimises  the JSD between  the target probability distribution $P$ and the joint probability distribution $p(\bs{\phi}\vert\theta)$ of the formulae $\bs{\phi}$. \guido{Notice that our approach requires the specification of a full joint probability distribution over the truth values of multiple formulae; should such a level of specification not be required, i.e. only some probability values need to be matched, the remaining values can be filled arbitrarily, compatibly with normalisation constraints.}

%******************************************************************************************************
%******************************************************************************************************
%******************************************************************************************************

\section{Methodology}
\label{sec:methodology}
%
%\red{Introduce here the appropriate  material to discuss methodology, expanding a little the text of QEST paper. Start from SMC, with more on bayesian SMC maybe, and them move to GP and GP-ucb. Justify this at the beginning, explaining that we cannot compute the true prob numerically - models are generally too big for this - so we rely on SMC methods, but then we face the problem of optimising an unknown function whose evaluations are noisy and computationally expensive. }

Solving the system design and  system identification problems requires us to optimise a function 
depending on  the joint probability distribution of the satisfaction of $d$-input formulae $\phi_1,\ldots,\phi_d$, which has to be computed for different values of the model parameters $\theta$. As numerical model checking algorithms for MiTL formulae suffer severely from state space explosion \cite{MC:Mereacre:2011:MTLmc}, we will revert to  \emph{statistical model checking} (SMC), which will be introduced in Section \ref{sec:smc}.
While SMC provides a feasible way to estimate the joint satisfaction probability, it remains a computationally intensive method, providing only \emph{noisy} estimates. 
%Furthermore, we need to consider $2^d$ truth combinations, some of which are likely to have a very small satisfaction probability. 
A possible solution, relying on the fact that estimation noise will be approximately Gaussian \guido{due to the Central Limit Theorem}, is to adopt a Bayesian viewpoint: we can treat the unknown function as a random function (arising from a suitable prior stochastic process) and  the numerical estimations based on SMC as (noisy) observations of the function value, which in turn enable a {\it posterior} prediction of the function values at new input points. This is the idea underlying {\it statistical emulation} \cite{Kennedy:Bayesian02}, and leads to a very elegant algorithm for optimisation. This framework will be introduced in Sections \ref{sec:GP} and \ref{sec:GP-UCB}. We will conclude discussing a first example based on the Poisson process, for which we can compare the numerical results against analytical formulae.

\subsection{Statistical model checking}
\label{sec:smc}

We briefly review the  estimation of  the probability of MiTL formulae by  Statistical Model Checking (SMC \cite{Younes2006,Younes2006b,Zuliani2009}).
Given a stochastic process with fixed parameters $\theta$,  a simulation algorithm is used to sample trajectories of the process. For each sampled trajectory, we run a model checking algorithm for MiTL (for instance, the offline monitoring procedure of~\cite{MC:Maler:FORMATS2004:MiTL}),  to establish whether $\phi$ is true or false, thus generating samples from a Bernoulli random variable $Z_\phi$, equal to 1 if and only if $\phi$ is true. 
SMC uses a statistical treatment of those samples, like Wald sequential testing \cite{Younes2006b} or Bayesian alternatives \cite{Zuliani2009}, to establish if the query $P(\phi\vert\theta) > q$ is true, with a chosen confidence level $\alpha$, given the evidence seen so far. 
Bayesian SMC, in particular, uses a Beta prior distribution $Beta(q\vert a,b)$ for the probability of $q=P(\phi=1)$; by exploiting the conjugacy of the Beta and Bernoulli distributions \cite{ML:Bishop:2006:PRandML}, applying Bayes' theorem we get 
\[  P(q\vert D_\phi) =  \frac{1}{P(D_\phi)} P(D_\phi\vert q) P(q) = Beta(q,a+k_1,b+k_0), \]
where $D_{\phi}$ is the simulated data, $k_1$ is the number of times $Z_\phi = 1$ and $k_0$ the number of observations of 0.
The parameters $a$ and $b$ of the Beta prior distribution (usually set to 1) can be seen as pseudo-counts that regularise the estimate when a truth value is rarely observed. Our best guess about the true probability $P(Z_\phi=\true)$ is then given by the predictive distribution \cite{ML:Bishop:2006:PRandML}: $P(Z_\phi=\true\vert D_\phi) = \bbE[q\vert D_\phi] = \frac{k_1+a}{k_1+a+k_0+b}$.
%\[ P(Z_\phi=\true\vert D_\phi) = \int_0^1 P(Z_\phi=\true\vert q) P(q\vert D_\phi)dq = \bbE[q\vert D_\phi] = \frac{k_1+a}{k_1+a+k_0+b}\]

The Bayesian approach to SMC, especially the use of prior distributions as a form of regularization of sampled truth values of formulae, is particularly relevant for our setting, since we need to estimate probabilities over $2^d$ joint truth values of $d$ formulae, i.e.  we need to sample from a discrete distribution $Z_{\phi_1,\ldots,\phi_d}$ with values in $\mathcal{D}=\{\true,\false\}^d$. Some of these truth combinations will be very unlikely, hence regularization is a crucial step to avoid errors caused by keeping reasonably small the number of runs.
In  order to extend Bayesian SMC to estimate the joint truth probabilities of $d$  formulae, \luca{we choose a Dirichlet prior distribution, which is a distribution on the unit simplex in $\bbR^n$, so that it can be seen as the multidimensional extension of the Beta distribution, and it also enjoys the conjugate prior property. The Dirichlet distribution has density \[Dirichlet(\vec q\vert \alpha_1,\ldots,\alpha_{2^d} ) \propto \prod_{i=1}^{2^d} q_i^{\alpha_i-1} \]} depending on $2^d$ parameters $\alpha_i$, which can be seen as pseudo-counts, and which we fix to one.\footnote{The corresponding Dirichlet distribution boils down to a uniform distribution.}
%
%
%(with $\alpha_0 = \sum_{j=1}{2^d}\alpha_j$, and $\Gamma$ being the Gamma function \cite{ML:Bishop:2006:PRandML}):
%\[ Dir(\vec q\vert \alpha_1,\ldots,\alpha_{2^d} ) = \frac{\Gamma(\alpha_0)}{\Gamma(\alpha_1)\cdots\Gamma(\alpha_2^d)} \prod_{j=1}^{2^d} q_j^{\alpha_j-1}.\] 
%The  parameters $\alpha_1,\ldots,\alpha_{2^d}$ of the Dirichlet prior are chosen equal to 1, corresponding to adding one pseudo-count to every possible joint truth value. 
%
Given observations $D_{\phi_1,\ldots,\phi_d}$ of the truth values of $Z_{\phi_1,\ldots,\phi_d}$\footnote{Note that $D_{\phi_1,\ldots,\phi_d}$ is a matrix, similarly the design matrix discussed in Section \ref{sec:problem}, but we treat each column/ observation as a single point of $\mathcal{D}$.}, analogous calculations yield  the posterior distribution over multinomial distributions on $\mathcal{D}$ as 
$ p(\vec q\vert D_{\phi_1,\ldots,\phi_d}) = Dirichlet(\vec q\vert \alpha_1+k_1,\ldots,\alpha_{2^d}+k_{2^d})$,
where $k_j$ is the number of times we observed the $j$th truth combination, corresponding to a point $\vec{d}_j\in\mathcal{D}$. Using the fact that the marginals of the Dirichlet distributions are Beta distributed,  the predictive distribution is readily computed as 
$p(Z_{\phi_1,\ldots,\phi_d} = \vec{d}_j\vert D_{\phi_1,\ldots,\phi_d}) = (\alpha_j+k_j)/(\alpha_0+k)$.
This probability is then used to estimate the likelihood $\mathcal{L}(D,\theta)$, as 
$ \mathcal{L}(D,\theta)=\prod_{i=1}^NP(D_i\vert\theta)$ or the JSD.
By the law of large numbers, with probability one, this quantity will converge to the true likelihood when the number of samples in the SMC procedure becomes large, and the deviation from the true likelihood will become approximately Gaussian.

%
%
%
%Point out that some combinations of truth values for certain parameters may be very unlikely, so we will need to regularise (add pseudocounts) or run many simulations. Usual limitations of statistical model checking. However emphasize that for large enough sample sizes the error is small and Gaussian distributed (as we are averaging; this is important below)

%%%%%%%%%%%%%%%%%%%%%%%%%%%%%%%%%%%%%%%%%%%%%%%%
%%%%%%%%%%%%%%%%%%%%%%%%%%%%%%%%%%%%%%%%%%%%%%%%
%%%%%%%%%%%%%%%%%%%%%%%%%%%%%%%%%%%%%%%%%%%%%%%%

%******************************************************************************************************
%******************************************************************************************************
%******************************************************************************************************
\subsection{Gaussian Process Regression}
\label{sec:GP}
A {\it Gaussian Process} (GP) is a probability measure over the space of continuous functions (over a suitable input space) such that all of its finite-dimensional marginals are multivariate normal. 
A GP is uniquely defined by its {\it mean} and {\it covariance} functions, denoted by $\mu(x)$ and $k(x,x')$. By definition, we  have  that for every finite set of points\begin{equation}
f\sim\mathcal{GP}(\mu,k)\leftrightarrow \mathbf{f}=\left(f(x_1),\ldots,f(x_N)\right)\sim\mathcal{N}\left(\boldsymbol{\mu},K\right)\label{GPdefinition}\end{equation}
where $\boldsymbol{\mu}$ is the vector obtained evaluating the mean function $\mu$ at every point, and $K$ is the matrix obtained by evaluating the covariance function $k$ at every pair of points. In the following, we will assume for simplicity that the prior mean function is identically zero (a non-zero mean can be added post-hoc to the predictions w.l.o.g.).

% Naturally, as the covariance of a Gaussian distribution must be a symmetric positive-definite matrix, equation (\ref{GPdefinition}) imposes some constraints on the allowable covariance functions; these are the so called {\it Mercer conditions} \cite{Rasmussen:Gaussian06}.

The choice of covariance function determines the type of functions which can be sampled from a GP (more precisely, it can assign prior probability zero to large subsets of the space of continuous functions). A popular choice of covariance function is the {\it radial basis function} (RBF) covariance\begin{equation}
k(x,x')=\gamma\exp\left[-\frac{\Vert x-x'\Vert^2}{\lambda^2}\right]\label{RBFcov}\end{equation}
which depends on two hyper-parameters, the {\it amplitude} $\gamma$ and the {\it lengthscale} $\lambda$. Sample functions from a GP with RBF covariance are with probability one infinitely differentiable functions. 
%******************************************************************************************************
%******************************************************************************************************
%******************************************************************************************************
GPs are a very natural framework for carrying out the {\it regression task}, i.e. estimating a function from observations of input-output pairs. Noisy observations of function values (a {\it training} set) can be combined with a GP prior to yield Bayesian posterior estimates of the function values at novel query input values. If the observation noise is Gaussian (as is the case we consider in this paper), the required computations can be performed analytically to yield a closed form for the predictive posterior.

%By using the basic rules of probability and matrix algebra,  that the joint probability of the observations and the function values at the input points $x_1,\ldots,x_N$ and the extra input value $x^*$ factorises as \begin{equation}
% p\left(\mathbf{y},f(x_1),\ldots,f(x_N),f(x^*)\right)=p\left(\mathbf{y}\vert f(x_1),\ldots,f(x_N)\right)p\left(f(x_1),\ldots,f(x_N),f(x^*)\right)\label{bigJoint}\end{equation}
% where the two terms on the r.h.s. are given by equation (\ref{likelihood}) and equation (\ref{GPdefinition}) respectively. Due to the assumption of independent Gaussian noise in the observations, it is trivial to integrate out (marginalise) the function values $f(x_1),\ldots,f(x_N)$ to obtain the following joint distribution for the observations and the function value at the new input point\begin{equation}
% p\left(\mathbf{y},f(x^*)\right)=\mathcal{N}(\boldsymbol{mu},\hat{K})\label{littleJoint}\end{equation}
% where $\hat{K}$ is obtained from $K$ in equation (\ref{GPdefinition}) by adding $\sigma^2I$ to the top left $N\times N$ block. Fixing the values of $\mathbf{y}$ to the observed ones (conditioning), it is now possible to compute analytically the {\it posterior} distribution over the function value at $x^*$ (sometimes also called {\it predictive} distribution) given the observations. As all the distributions involved are Gaussian, this posterior will also be Gaussian\begin{equation}
% p\left(f(x^*)\vert\mathbf{y}\right)=\mathcal{N}\left(\mu^*,k*\right).\label{GPpost}\end{equation}
 Assuming for simplicity a zero prior mean function,
we have that the predictive distribution at a new input $x^*$ is Gaussian with mean \begin{equation}
\mu^*=\left(k(x^*,x_1),\ldots,k(x^*,x_N)\right)\hat{K}_N^{-1}\mathbf{y}\label{GPpostMean}\end{equation}
\vspace{-0.2cm}
and variance\begin{equation}
k^*=k(x^*,x^*)-\left(k(x^*,x_1),\ldots,k(x^*,x_N)\right)\hat{K}_N^{-1}\left(k(x^*,x_1),\ldots,k(x^*,x_N)\right)^T.\label{GPpostVar}\end{equation}
\guido{Here, $\mathbf{y}$ is the vector of observation values at the training points $x_1,\ldots,x_N$ and \[
\hat{K}_N(i,j)=k(x_i,x_j)+\delta_{ij}\sigma^2_i\] 
with $\sigma^2_i$  the observation noise variance at point $x_i$ (see below section \ref{sec:heteroschedastic} for how this quantity is estimated in our case).} Notice that the first term on the r.h.s of equation (\ref{GPpostVar}) is the prior variance at the new input point; therefore, we see that the observations lead to a {\it reduction} of the uncertainty over the function value at the new point. The variance however returns to the prior variance when the new point becomes very far from the observation points.

GPs are a rich and dynamic field of research in statistical machine learning, and this quick introduction cannot do justice to the field. For more details, we refer the interested reader to the excellent review book of Rasmussen and Williams \cite{Rasmussen:Gaussian06}.

%******************************************************************************************************
%******************************************************************************************************
%******************************************************************************************************
\subsection{Bayesian optimisation}
\label{sec:GP-UCB}

We now return to the problem of finding the maximum of an unknown function with the minimum possible number of function evaluations. The underlying idea of {\it Bayesian Optimisation} (BO) is to use a probabilistic model  (e.g. a GP) to estimate (with uncertainty) a statistical surrogate of the unknown function (this is sometimes called emulation in the statistics literature \cite{Kennedy:Bayesian02}). This allows us to recast the optimisation problem in terms of
trade off between the {\it exploitation} of promising regions (where the surrogate function takes high values) with the {\it exploration} of new regions (where the surrogate function is very uncertain, and hence high values may be hidden).

%\red{POSSIBILE TAGLIARE DA QUI} 
%This is related to the problem of performing sensitivity analysis w.r.t. the parameters of complex computer models, e.g. climate models, where a quantification of uncertainty on the model outputs is essential. An elegant approach to solving this problem has been proposed by Kennedy and O'Hagan \cite{Kennedy:Bayesian02} by recasting the problem in a Bayesian formalism: the true function linking the parameters to the model outputs is assumed unknown and is assigned a GP prior. A (limited) number of function evaluation are then used as (noiseless) observations to obtain a GP posterior mean function which {\it emulates} the true unknown function, and is used for subsequent analyses.

%In the optimisation case, the situation is slightly different: given an initial set of function evaluations, we are interested in determining a sequence of input values that converges to the optimal value of the function. %\red{A QUI} 
%A naive approach would be to use GP regression to emulate the unknown function, and to explore the region near the maximum of the posterior mean. It is easy to see, though, that this approach is vulnerable to remaining trapped in local optima. On the other hand, one could sample uniformly across the input domain of interest; this is guaranteed to eventually find the global optimum but is unlikely to do so in a reasonable time. It is therefore clear that one needs to 

Optimal trade-off of exploration and exploitation is a central problem in reinforcement learning, and has attracted considerable theoretical and applicative research. Here we use the GP Upper Confidence Bound (GP-UCB) algorithm \cite{Srinivas:information12}, an exploration-exploitation trade-off strategy which provably converges to the global optimum of the function. The idea is intuitively very simple: rather than maximising the posterior mean function, one maximises an upper quantile of the distribution, obtained as mean value plus a constant times the standard deviation (e.g. the 95\% quantile, approximately given as $\mu+2\sigma$). The GP-UCB rule is therefore defined as follows: let $\mu_t(x)$ and $var_t(x)$ be the GP posterior mean and variance at $x$ after $t$ iterations of the algorithm. The next input point is then selected as\begin{equation}
x_{t+1}=\mathrm{arg max}_x\left[\mu_t(x)+\beta_t\sqrt{var_t(x)}\right]\label{GPUCBrule}
\end{equation}
where $\beta_t$ is a constant that depends on the iteration of the algorithm. The importance of the work of \cite{Srinivas:information12} lies in the first proof of convergence for such an algorithm: they showed that, with high probability,   the algorithm is {\it no-regret}, i.e.  \vspace{-0.1cm} \[
\lim_{T\rightarrow\infty}\frac{1}{T}\sum_{t=1}^T\left((x^*)-f(x_t)\right)=0.\] 
where $x^*$ is the true optimum and $x_t$ is the point selected with the UCB rule at iteration $t$.

\vspace{0.2cm}

\noindent{\bf Remark}: \guido{The use of a GP with RBF covariance implicitly limits the set of possible emulating functions to (a subset of the set of) smooth functions. This is not a problem when optimising parameters of a CTMC: it was recently proved in \cite{Bortolussi:smoothed14} that the satisfaction probability of a MiTL formula over a CTMC is a smooth function of the model parameters, which immediately implies that the likelihood of truth observations is also smooth. In general, we conjecture that smoothness will hold for purely stochastic processes, i.e. systems where it is impossible to find a (strict) subset of state variables $X_J$ such that the system dynamics conditioned on $X_J$ are deterministic. It is easy to show that smoothness does not hold for deterministic systems, where the satisfaction probability can jump from zero to one as the parameters are varied. In hybrid deterministic/ stochastic processes, smoothness may therefore not hold; in these cases, the algorithm will still execute, but its convergence guarantees will be lost, so that application of our method to this class of systems should be considered as heuristic.}

\vspace{-0.1cm}
%Srinivas et al \cite{Srinivas:information12} then proved the following theorem
%\begin{theorem}
%Let $\beta_t=k+\alpha\log t$, where $k$ and $\alpha$ are positive constants. Then the GP-UCB algorithm in equation (\ref{GPUCBrule}) is no-regret. More specifically, with high probability, the cumulative regret is bounded by $O(\sqrt{T})$.
%\end{theorem}
%This theorem indicates that, as the algorithm proceeds, exploration needs to become gradually more important than exploitation ($\beta_t$ is monotonically increasing), as one would intuitively expect. The algorithm has been successfully employed in a number of difficult optimisation problems, from determining optimal structure of synthetic proteins \cite{Romero:navigating13} to computer vision \cite{vezhnevets:weakly12}.

\subsection{Example: Poisson process}
\label{sec:poisson}

\luca{As a simple example illustrating our approach, we consider observing the truth values of an atomic proposition over realisations of a Poisson process.  We briefly recall that a Poisson process with rate $\mu$ is an increasing, integer valued process such that
\begin{equation}
P(k=n\vert\mu, t)=\frac{(\mu t)^n}{n!}\exp[-\mu t].\label{PoissMarg}
\end{equation}
Poisson processes are fundamental in many applications, ranging from molecular biology to queueing theory, where they often form the basic building blocks of more complex models.
We consider a very simple scenario where we have observed the truth value of the formula \[\phi= \mathbf{F}^{[0,1]} \{k>3\},\] i.e. the formula expressing the fact that $k$ has become bigger than 3 within 1 time unit, evaluated on individual trajectories sampled from a process with $\mu=2$. The probability of  
$\phi$ being true for a trajectory given the value of $\mu$ can be calculated analytically as\begin{equation}
p(\mu) =P(\phi=true)=1-P(\phi=false)=1-\sum_{n=0}^3\frac{(\mu)^n}{n!}\exp[-\mu].\label{analVals}\end{equation}
This leads to the following analytical formula for the log-likelihood, given a fixed set of observations $D$ of its truth:
\begin{equation}
 \mathcal{L}(\mu,D) =  \#_{true}(D) \log(p(\mu)) + \#_{false}(D)\log( 1- p(\mu) ), \label{Poissloglike}
\end{equation}
where $\#_{true}(D)$ counts the number of times the formula was observed true in $D$, and $\#_{false}(D)$ counts the occurences of $false$ in $D$. This gives us an ideal benchmark for our approach.}

\begin{figure}[!t]
\begin{center}
\subfigure[]{\includegraphics[width=0.3\textwidth]{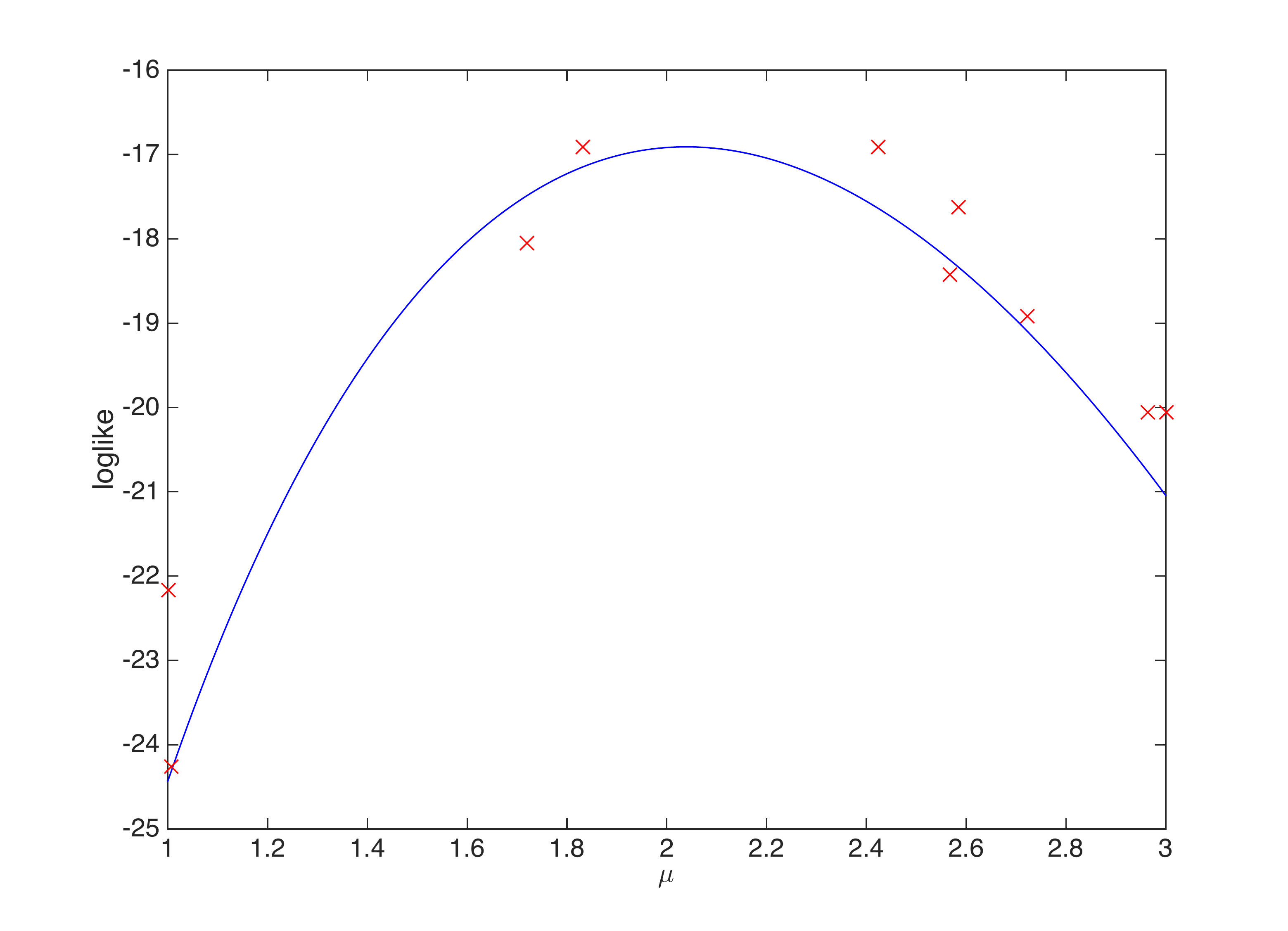} \label{poissona}}
\subfigure[]{\includegraphics[width=0.3\textwidth]{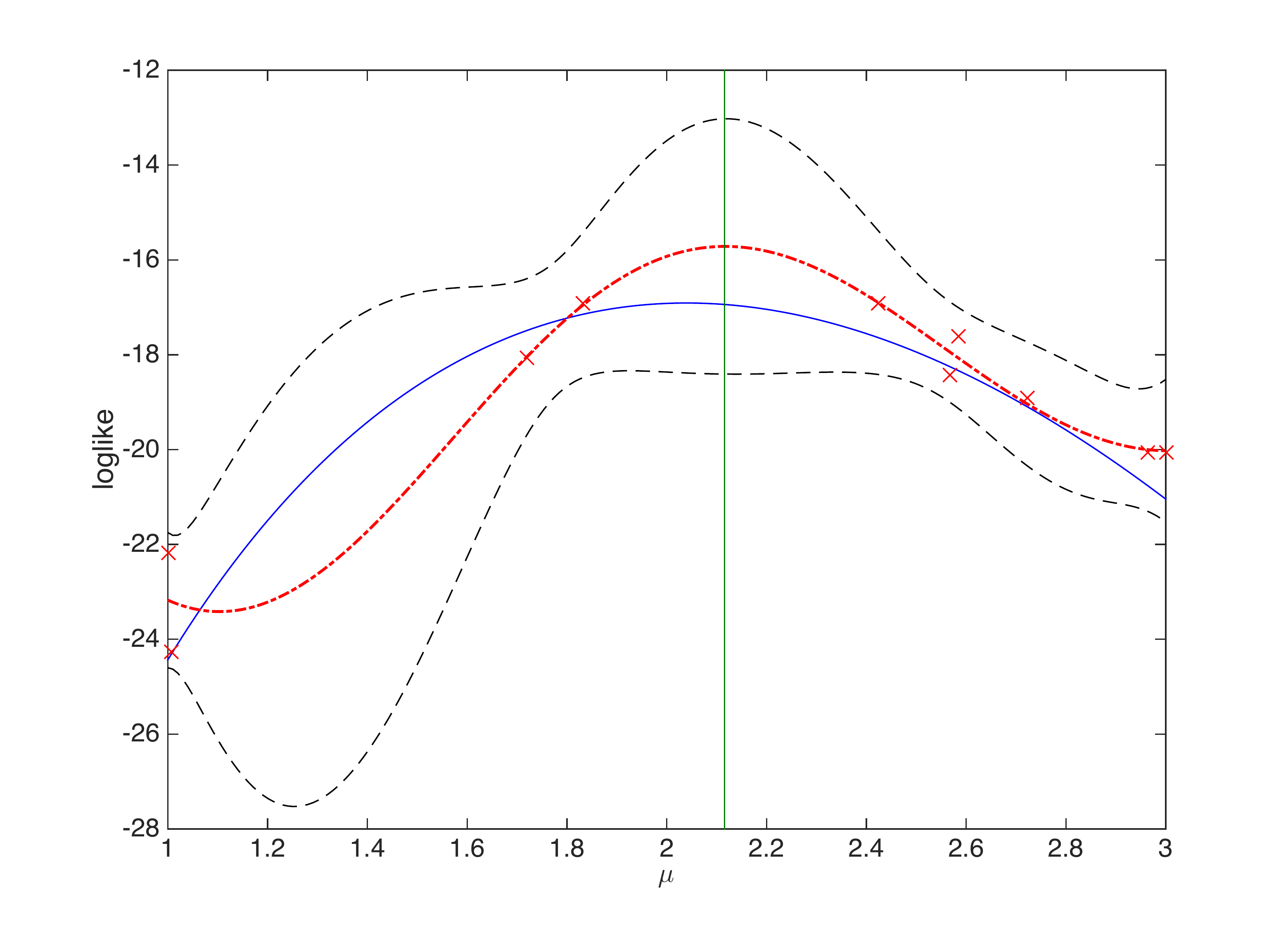} \label{poissonb}}
\subfigure[]{\includegraphics[width=0.3\textwidth]{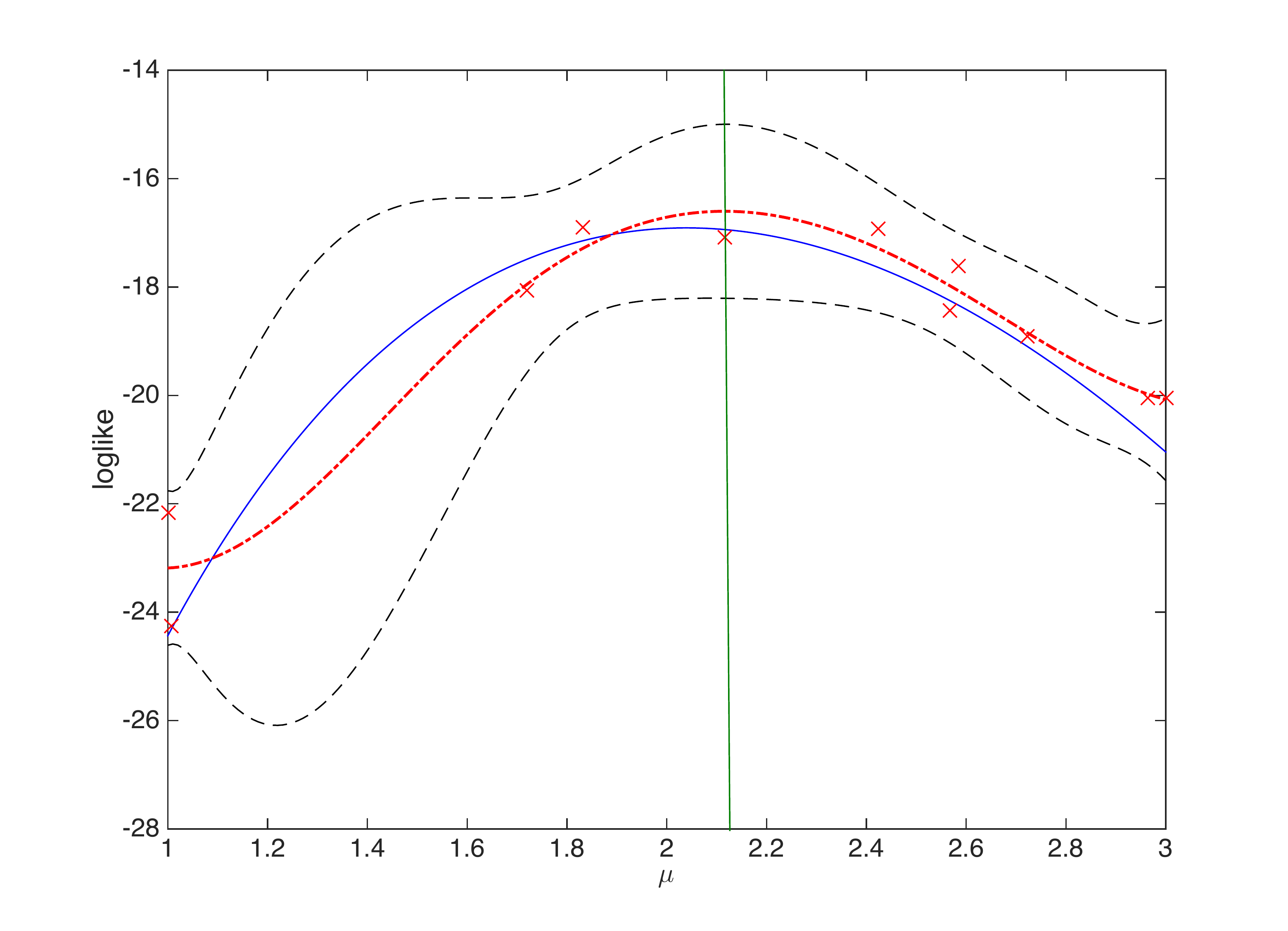} \label{poissonc}}

\subfigure[]{\includegraphics[width=0.3\textwidth]{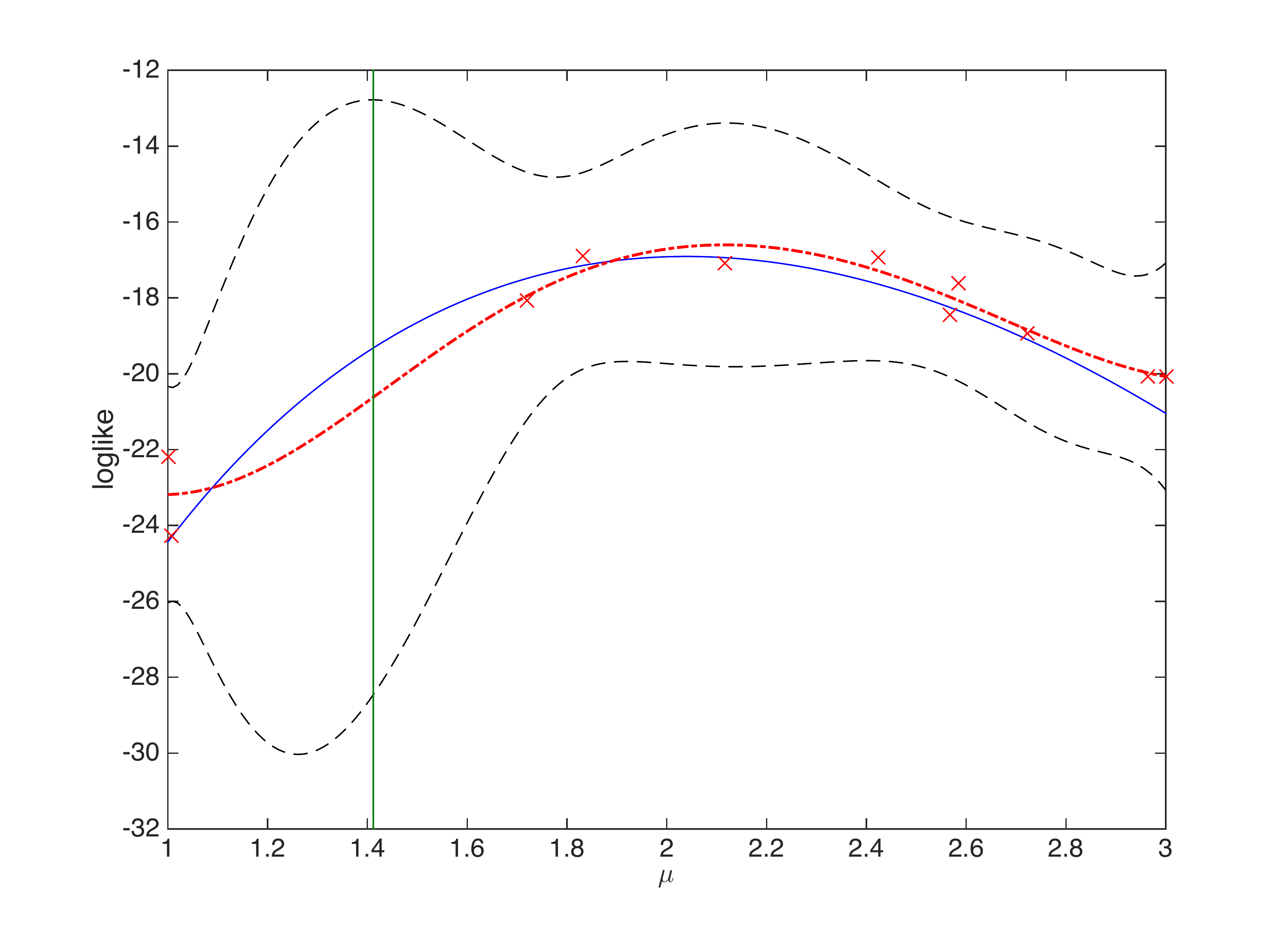} \label{poissond}}
\subfigure[]{\includegraphics[width=0.3\textwidth]{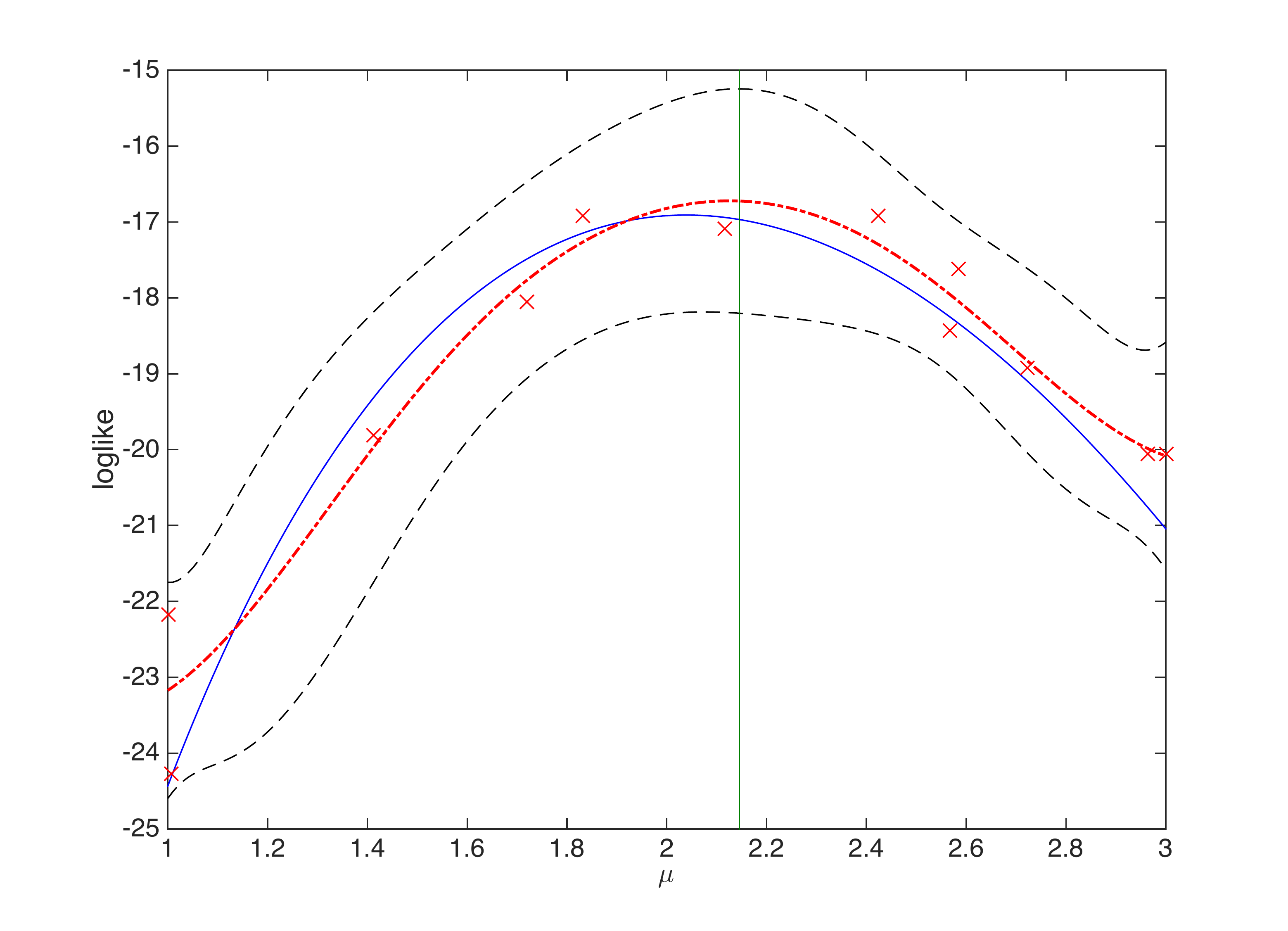} \label{poissone}}
\end{center}

\caption{\luca{Illustration of the learning procedure on the Poisson process example.  ({\bf\ref{poissona}}) Exact log likelihood (computed from formula (\ref{Poissloglike})) and 10 SMC estimation (from 100 simulation runs, red crosses) for $\mu\in[1,3]$.  ({\bf\ref{poissonb}}) Illustration of the GP-UCB algorithm: GP likelihood estimation (red dash-dotted line), true likelihood (solid blue line), and GP-UCB confidence bounds (black dotted lines). The maximum of the upper confidence bound is identified by the vertical line. ({\bf\ref{poissonc}}) Illustration of the GP-UCB algorithm, after sampling the log-likelihood (by SMC estimation) at the maximum value previously identified. Lines have the same meaning as before. Notice the reduced variance near the new sampled point. The maximum, however is predicted at the same place. ({\bf\ref{poissond}}) Illustration of the GP-UCB algorithm after increasing $\beta_t$ from 2 to 4. Now the maximum of the upper confidence bound is in an area of high uncertainty, on the left. ({\bf\ref{poissone}}) Illustration of the GP-UCB algorithm, after sampling the log-likelihood at the maximum identified in \ref{poissond}. Again, notice how the uncertainty is reduced.}
}\label{figurinaPoisson}

%\vspace{-0.5cm}

\end{figure}

\luca{Figure~\ref{figurinaPoisson} shows a generic step of the GP-UCB algorithm at work. The starting point is the set $D$, in this case containing 40  independent observations of process trajectories. The exact log likelihood is computed according to equation (\ref{Poissloglike}), and shown in Figure~\ref{poissona}, together with 10 samples of the log-likelihood, computed by SMC (red dots).  In  Figure~\ref{poissonb}, we show the result of running GP-regression over these 10 observations. The predictive posterior mean is in red, while the dashed black lines represent the upper and  lower confidence bounds of the distribution on functions defined by the posterior GP,  for $\beta_t\equiv2$. The vertical line is the maximum identified by the global search procedure needed to optimise the GP upper confidence bound. The log-likelihood is then sampled at this new point, again by SMC, and the GP regression is run again on such an enlarged input set. The result is shown in Figure~\ref{poissonc}. As can be seen, the variance of the  prediction, i.e. the width of the upper confidence bound, has been considerably reduced in the region around the new input point for the GP regression task. In this case, however, the maximum of the upper confidence bound is not changed, hence we increase  $\beta_t$ from 2 to 4. The result is shown in Figure~\ref{poissond}, where we can see that the maximum is now shifted on the right, on a high uncertainty region. The log-likelihood is sampled again at this newly identified point, and the result is a large reduction of variance for small $\mu$, as can be seen in Figure~\ref{poissone}. }

%******************************************************************************************************
%******************************************************************************************************
%******************************************************************************************************

%%%%%%%%%%%%%%%%%%%%%%%%%%%%%%%%%%%%%%%%%%%%%%%%
%%%%%%%%%%%%%%%%%%%%%%%%%%%%%%%%%%%%%%%%%%%%%%%%
%%%%%%%%%%%%%%%%%%%%%%%%%%%%%%%%%%%%%%%%%%%%%%%%

\section{Enhancing the methodology}
\label{sec:enhancements}

We \guido{briefly} discuss here some further improvements to the basic methodology presented in Section \ref{sec:methodology}; \guido{an extensive tutorial introduction of the statistical concepts used is beyond the scope of this paper, however full details can be found in the referenced literature. The methodological enhancements are discussed} in the context of system identification using the likelihood, but similar considerations apply for design or MAP identification. 
%
%\red{Discuss weak points of the previous approach and possible solutions. We do not have an estimate of the intrinsic reliability of the prediction made by the algorithm. A way to tackle this problem is the so called Laplace approximation. Another possibility is the variational approximation of the posterior.\\
%Observation noise depends on the parameters: it is heteroschedastic. Discuss how to incorporate this on the algorithm and how to estimate such noise: with posterior or bootstrapping?\\
%Lengthscale: it influences the performance of the method, now it is black magic. Use a model selection strategy optimising the marginal likelihood (to be done).  }

\subsection{Laplace approximation}
\label{sec:laplace}

%\red{TO BE DONE, JUST CUT AND PASTED FROM QEST}
The GP-UCB algorithm enables us to find the maximum of a function (in our case, the likelihood function or the un-normalised posterior); in many cases, however, it is very desirable to be able to provide uncertainty estimates over the parameter values returned. Given the intractable nature of the likelihood, which requires a computationally expensive statistical model checking procedure at every parameter value, a fully Bayesian treatment (e.g. based on Markov chain Monte Carlo simulations) is ruled out. A cheaper alternative is to compute a local Gaussian approximation: this procedure, known as {\it Laplace approximation} in statistics and machine learning, approximates the uncertainty as the inverse of the local curvature at the optimum (see e.g. \cite{ML:Bishop:2006:PRandML}, Ch 4.4). This is equivalent to locally approximating the log-likelihood with a quadratic form. In our case, we cannot directly compute derivatives of the unknown likelihood function: instead, we compute a Laplace approximation to the GP mean function at the optimum. In practice, as the GP is only estimated on a discrete subset of points, we perform a local optimisation step around the GP-UCB solution (by using the Newton-Raphson algorithm \guido{applied to the posterior GP mean function}), and compute the Hessian at the resulting maximum. \guido{The inverse of the Hessian matrix then provides a local approximation to the covariance structure of the estimated parameters: in particular, its diagonal entries can be used to provide confidence values over the estimated parameters.} Naturally, the local properties of the GP posterior mean are influenced both by the true underlying function, but also on the hyper-parameters of the GP prior; we discuss below how such hyperparameters can be set automatically using an additional optimisation step.

\subsection{Heteroschedastic noise} 
\label{sec:heteroschedastic}

One drawback of the method introduced in Section~\ref{sec:methodology} is that it assumes the observation noise to be uniform in the whole parameter space. This is an oversimplification, as the noise in the SMC estimation of the likelihood is heteroschedastic, i.e. it depends  on the joint satisfaction probability and on the variability of the estimate. A non-homogeneous treatment of noise will reduce the variability in the estimation of the likelihood function.\footnote{Notice that the only difference in GP regression is that now the covariance matrix of observation added to $\hat{K}_N$ is diagonal with non-constant elements on the diagonal.} We propose two approaches to compute the noise in the log-likelihood, the first one computational, based on bootstrapping, and the second one analytic, exploiting the nature of the posterior distribution of Bayesian SMC. 

% Intuitively, a joint satisfaction probability which puts mass one in a single point will be estimated less noisily than a uniform satisfaction probability. If we can compute with higher precision the likelihood at a certain parameter value, we should take this into account, as it will reduce the variability in the estimation of the likelihood function.

%Technically, if we have an estimate $\sigma(\vec{x})$ of the standard deviation of the log-likelihood function at each input value $\vec{x_i}$ we can include it in the GP regression by properly defining the matrix $\hat{K}_N$.  
%In particular, we just need to replace the matrix $\sigma I$ that is added to the Gram matrix $K$ in equation XXX by the matrix $diag(\sigma(\vec{x_1},\ldots,\vec{x_n}))$. This will assign a specific noise level to each input point. 

%The more challenging aspect is the estimation of the standard deviation of the log-likelihood. We propose here two approaches, the first one computational, based on bootstrapping, and the second one analytic, exploiting the nature of the posterior distribution of the Bayesian SMC routine we use. 

\subsubsection{Bootstrapping} 
Bootstrapping is a standard statistical technique to obtain estimates of confidence intervals \cite{ML:Bishop:2006:PRandML}. In our setting, it works by resampling with repetition from the set of observed joint truth values, and recomputing the log-likelihood from each sampled set. In this way, one obtains an empirical distribution of the log-likelihood, from which statistics and confidence intervals can be extracted. In our case, the bootstrapped statistic is the standard deviation of the empirical bootstrap distribution.

\subsubsection{Posterior estimate} 
As an alternative to bootstrapping, we can exploit the fact that Bayesian SMC gives us a posterior distribution on  the space of probability distributions over the joint truth value  of $d$ MiTL formulae $\phi_1,\ldots,\phi_d$.
We recall that, assuming a Dirichlet prior with $2^d$ parameters $\boldsymbol{\alpha} = (\alpha_i)$,  and if $k_j$ simulations resulted in the truth value $\vec{d}_j$, then the posterior distribution is again Dirichlet  with parameters $\boldsymbol{\alpha} + \vec{k}$.

A typical Bayesian treatment of noise is to compute the average distribution of the quantity of interest with respect to the posterior distribution, thus taking into account the full noise distribution. Recall that the likelihood can be seen as a function of $\vec{q}$, with $\vec{q}\sim Dirichlet(\boldsymbol{\alpha} + \vec{k})$, and let $\vec{h}$ be the vector counting truth value in the observations $D$, $h_j = \#(\vec{d}_j,D)$. Simple computations give $\bbE[L(\vec{q})] = \frac{B(\boldsymbol{\alpha} + \vec{k} + \vec{h})}{B(\boldsymbol{\alpha} + \vec{k})}$ and $VAR[L(\vec{q}) ]  =  \frac{B(\boldsymbol{\alpha} + \vec{k} + 2\vec{h})}{B(\boldsymbol{\alpha} + \vec{k})} -  \frac{B^2(\boldsymbol{\alpha} + \vec{k} + \vec{h})}{B^2(\boldsymbol{\alpha} + \vec{k})}$, where $B(\vec{x}) = \frac{\prod_{i=1}^{2^d} \Gamma(x_i) }{\Gamma(\sum_{i=1}^{2^d}x_i)}$ is the multinomial Beta function and $\Gamma(x) = \int_0^\infty y^{x-1}e^{-y}dy$ is the gamma function.

\subsection{Optimisation of hyperparameters}
\label{sec:hyperparameters}
A delicate issue about GP-UCB optimisation is that the emulation of the log-likelihood or of the JSD depends on the choice of hyperparameters of the kernel, which in the case of the RBF Gaussian kernel are the amplitude $\alpha$ and the lengthscale $\lambda$. The problem of leaving this choice to the user is that the results of the optimisation and its computational complexity (number of log-likelihood evaluations) depend in unpredictable ways on these parameters, particularly on the lengthscale. 
%Finding the correct value requires either a lot of patience or a good intuition about the problem. 
In fact, the lengthscale governs the Lipschitz constant of the  functions sampled from the GP, hence of the posterior prediction, especially at a low number of input points. It would be wise, therefore to try to estimate such hyperparameters from the  batch of  initial observations. There are two main approaches to do this~\cite{Rasmussen:Gaussian06}. One way is to take a Bayesian perspective and put a prior on hyperparameters, estimating their posterior distribution from observed data via Monte Carlo sampling, which is unfeasible in our setting.  Alternatively, we can treat the estimation of hyperparameters as a model selection problem, which can be tackled in a maximum-likelihood perspective by optimising the model evidence
\[p(\vec{y}|X,\alpha,\lambda) = \int p(\vec{y}|\vec{f},X)p(\vec{f}|X,\alpha,\lambda) d\vec{f}. \]
Essentially, $p(\vec{y}|X,\alpha,\lambda)$ is the marginal likelihood of the observed data, computed by marginalising the product of the likelihood times the GP prior, and is a function of the hyperparameters for which an analytic expression can be derived \cite{Rasmussen:Gaussian06}.
The idea behind is that the larger this value, the better the hyperparameters, and hence the GP, explain the observed data. 
%
%Fixing the hyperparameters using the model evidence has two advantages: first, we retain the analytic treatment of GP regression and, secondly, there is an analytic expression of the log-marginal likelihood, which can be easily computed as
%\[ \log p(\vec{y}|X,\alpha,\lambda) = -\frac{1}{2}\vec{y}^T(K+\sigma^2 I)^{-1}\vec{y} - \frac{1}{2}\log|K+\sigma^2 I | - \frac{n}{2}\log 2\pi,  \]
%where $K$ is the Gram matrix, $\sigma^2$ is the variance of the observation noise\footnote{In case of heteroschedastic noise, we just need to replace $\sigma^2 I$ with the matrix $diag(\sigma^2_1,\ldots,\sigma^2_n)$, where $\sigma^2_1,\ldots,\sigma^2_n$ is are the estimated variance of observations at input points, computed with one of the methods of Section \ref{sec:heteroschedastic}. }, and $|K+\sigma^2 I |$ is the determinant of the matrix $K+\sigma^2 I$.
In this paper, we take this second approach. As the model evidence can potentially have multiple local maxima, we use a simple global optimisation scheme, running several times a Newton-Raphson local optimisation algorithm from random starting points \cite{Burden2011}. Experimentally, we found that the model evidence tends to behave quite well, with a global optimum having a large basin of attraction, a phenomenon often observed in practice~\cite{Rasmussen:Gaussian06}, so that few runs, on the order of five, of the optimisation routine suffice. 

%\red{Should we say that we use the matlab optimisation toolbox? Does ED university has a licence of this toolbox, when one logs to ease???}

\subsection{Grid sampling strategies}
\label{sec:sampling}
A final improvement of the algorithm we consider in this paper is related to the sampling strategies for the initial set of points at which the likelihood or the JSD is evaluated, and the sampling strategy for the  points at which the emulated likelihood is computed to look for a maximum. The goal is to maximise the coverage of the parameter space, keeping the number of sampled points to a minimum. Simple but effective schemes in this respect are based on the \emph{latin hypercube sampling} strategy (LHS) \cite{mckay1992latin}, which splits a $d$-dimensional cube into $k^d$ smaller cubes, and samples $k$ points, at most one for each smaller cube, with the constraint that two sampled points cannot belong to cubes that overlap when projected in any of the $d$ dimensions. For $d=2$, LHS samples a latin square, from which it derives the name. The sampling approach we use  is a variation of LHS, called \emph{orthogonal LHS}, which further subdivides the space into equally probable subspaces. Points  sampled still satisfy the LHS property, with the further constraint that each subspace contains the same number of points, thus improving the coverage \cite{tang1993orthogonal}.

%%%%%%%%%%%%%%%%%%%%%%%%%%%%%%%%%%%%%%%%%%%%%%%%
%%%%%%%%%%%%%%%%%%%%%%%%%%%%%%%%%%%%%%%%%%%%%%%%
%%%%%%%%%%%%%%%%%%%%%%%%%%%%%%%%%%%%%%%%%%%%%%%%

\section{Experiments}
\label{sec:experiments}
In this section we will discuss two examples in more detail:  a simple CTMC model of rumour spreading in a social network, which resembles the diffusion of an epidemics, and a more complex SHS model of the toggle-switch, a simple genetic network composed of two genes repressing each other that shows bistable behaviour.

%
%\red{Here we discuss experimental results. We can use the network epidemics model, possibly enhanced with different classes of nodes (to mimick real networks), and a stochastic hybrid model of a toggle switch. Should we anticipate part of this discussion before presenting the enhancements, for instance using the Poisson process to illustrate the method? Should we discuss only one model (in this case the SHS)?  }
%

\subsection{Rumour spreading}
\label{sec:example:rumour}
The spreading or rumours or information in a social network is a phenomenon that has  received a lot of attention  since the sixties. %In particular, this phenomenon has many similarities with epidemic spreading, and in fact many proposed models  owe a lot to epidemiology, see \cite{Barrat2008}.
Here we consider a simple model \cite{Daley1964}, in which agents are divided into three classes: those that have not heard the rumour, the \emph{ignorants} ($I$), those that have heard the rumour and are actively spreading it, the \emph{spreaders} ($S$), and those that have stopped spreading it, the \emph{repressors} ($R$). The dynamics is given by three simple rules: when an ignorant comes into contact with a spreader,  the rumour is transmitted at rate $k_s$, while when a spreader comes into contact with another spreader or with a repressor, it stops spreading the rumour. This happens at rate $k_r$. We further multiply those rates by the average degree of connectivity $\langle k \rangle$ in the social network, i.e. the average number of people one is in contact with. The use of the average degree corresponds to the hypothesis of a homogeneous social network, see \cite{Barrat2008}. 
Summarising, the model  is a CTMC on three populations, $V_I$, $V_S$, and  $V_R$, subject to three types of events, which in the reaction-rate style \cite{SB:Gillespie:1977:gillespieAlgorithm} are:
\begin{itemize}
\item $V_I + V_S \rightarrow V_S + V_S$, with rate function $a_s(\vec{V},k_s,\langle k \rangle) = \frac{k_s \langle k \rangle}{N}\cdot V_S\cdot V_I$;
\item $V_S + V_S \rightarrow V_R + V_S$, with rate function $a_{r1}(\vec{V},k_r,\langle k \rangle) = \frac{k_r \langle k \rangle}{N}\cdot V_S\cdot V_S$;
\item $V_S + V_R \rightarrow V_R + V_R$, with rate function $a_{r2}(\vec{V},k_r,\langle k \rangle) = \frac{k_r \langle k \rangle}{N}\cdot V_S\cdot V_R$;
\end{itemize}
Notice the normalisation factor $N$ (the total population), which corresponds to a density dependence assumption, i.e. to a constant rate of contact per person, which is then multiplied by the probability of finding a spreader or a repressor, assuming random neighbours, see \cite{Daley1964,Barrat2008}. 
%The non-linearity of the passage from spreader to repressor is what distinguishes this model from an epidemiological one, where repressors correspond to recovered individuals and agents recover spontaneously,  without external interaction.
%
%
%--------------------------------------------------------------------------------------------------------------------------
%
For this system, we considered four temporal logic properties, expressed as MiTL formulae,  concerned with the number of spreaders and repressors, fixing the total population to  100. The properties are:
\begin{enumerate}
\item $\always{0}{200} (V_S < 45)$: the fraction of spreaders never exceeds 45\% in the first 200 time units. This  bounds the number of active spreaders from above;
\item $\eventually{22}{40}  (V_S > 35)$: between time 22 and 40, the fraction of spreaders exceeds 35\%. This locates the spreading peak between time 22 and 40;
\item $(\eventually{65}{90} (V_s = 0)) \wedge (\always{0}{65} (V_S > 0))$: the spreading process stops between time 65 and 90, and is active before time 65.
\item  $\always{90}{200} (82 < V_R < 88)$: the  fraction of repressors stabilises  from time 90 to 200 at between 82\% and 88\%. This corresponds to the fraction of population having heard the rumour. 
\end{enumerate}

%--------------------------------------------------------------------------------------------------------------------------

\subsubsection{Experimental Setup}  We fixed the average degree $\langle k \rangle$ to 20,\footnote{Note that the average degree multiplies all rates, hence fixing it corresponds to fixing the time scale.} while the remaining parameters are explored. To test the method under different conditions, we sampled uniformly $k_s \in [0.8,1.2]$ and $k_r\in[0.6,1.0]$, and use the sampled configuration to generate 40 observations $D$ of the value of the logical formulae. Then, we ran 20 times the GP-UCB optimisation algorithm in the following search space: $k_s \in [0.1,10]$, $k_r \in [0.08,8]$, so that each parameter domain spans over two orders of magnitude.
To treat equally each order of magnitude, as customary we transformed logarithmically the search space, and rescaled each coordinate into $[-1,1]$ (log-normalisation). 
The algorithm first computes the likelihood, using statistical model checking, for  48 points sampled randomly according to the orthogonal LHS strategy from the log-normalized space, and then uses the GP-UCB algorithm to estimate the position of a potential maximum of the upper bound function in a grid of 500 points, again sampled using orthogonal LHS. Noise is treated heteroschedastically, using bootstrapping, and the other hyperparameters are optimised after the computation of the likelihood for the initial points. 
If in the larger grid a point is found with  value greater than those of the observation points, we run a local optimisation algorithm (a Newton-Raphson scheme) to find the exact local maximum nearby, and then  compute the likelihood for this point and add it to the observations (thus changing the GP approximation). 
Termination happens when no improvement can be made after three grid resamplings. The algorithm terminated  after  only 10-15 additional likelihood evaluations on average. 

Results are reported for ML and MAP, in this case using  independent, vaguely informative Gamma priors, with mean $1$ for $k_s$ and $0.8$ for $k_r$, and shape equal to $10$. 
We also compare the effect of different enhancements, fixing the ``true'' parameter values to $k_s = 1.0$ and $k_r =0.8$ and the 40 observations, and running 100 times the algorithm for each combination of features.

\begin{table}
\[
\begin{array}{|c|c|c|c|c|c|c|c|}
\hline
\text{true $k_s$}  & \text{mean $k_s$} & \text{median $k_s$} & \text{std dev $k_s$} & \text{true $k_r$} & \text{mean $k_r$} & \text{median $k_r$} & \text{std dev $k_r$}\\
\hline
1.0313 & 1.048 & 1.0433 & 0.0714 & 0.6284 & 0.6308 & 0.626 & 0.0215\\
\hline
1.1674 & 1.2544 & 1.2536 & 0.0497 & 0.7481 & 0.7535 & 0.7565 & 0.0261\\
\hline
1.0806 & 1.0794 & 1.1052 & 0.1203 & 0.7775 & 0.813 & 0.834 & 0.0622\\
\hline
0.8112 & 0.7817 & 0.8071 & 0.1049 & 0.8332 & 1.0202 & 0.9075 & 0.3017\\
\hline
1.1231 & 1.0344 & 1.0357 & 0.041 & 0.9894 & 1.0086 & 1.0083 & 0.0385\\
\hline
0.8888 & 0.8265 & 0.8551 & 0.0794 & 0.9818 & 1.1193 & 1.0057 & 0.2257\\
\hline
1.0125 & 1.0382 & 1.0304 & 0.0459 & 0.6957 & 0.7363 & 0.7343 & 0.0271\\
\hline
1.0338 & 1.0422 & 1.0511 & 0.0469 & 0.6109 & 0.6052 & 0.6044 & 0.0273\\
\hline
0.9312 & 0.9096 & 0.9053 & 0.0296 & 0.7596 & 0.791 & 0.7831 & 0.0334\\
\hline
0.8606 & 0.7079 & 0.7083 & 0.081 & 0.8692 & 1.2647 & 1.1302 & 0.3749\\
\hline
\end{array} 
\]
\caption{Results for the maximum likelihood learning problem for the rumours spreading model. We report mean, median, and standard deviation on 20 runs, for 10 different true parameter combinations. }
\label{table:rumoursML}
\end{table}

\begin{table}
\[
\begin{array}{|c|c|c|c|c|c|c|c|}
\hline
\text{true $k_s$}  & \text{mean $k_s$} & \text{median $k_s$} & \text{std dev $k_s$} & \text{true $k_r$} & \text{mean $k_r$} & \text{median $k_r$} & \text{std dev $k_r$}\\
\hline
0.8321 & 0.7897 & 0.7963 & 0.0497 & 0.9747 & 1.1186 & 1.0952 & 0.0946\\
\hline
1.1046 & 1.125 & 1.1325 & 0.0407 & 0.6566 & 0.6771 & 0.6758 & 0.0234\\
\hline
1.1541 & 1.1698 & 1.167 & 0.0412 & 0.6934 & 0.6944 & 0.699 & 0.0265\\
\hline
0.938 & 1.0129 & 1.0182 & 0.0409 & 0.7254 & 0.7618 & 0.7616 & 0.0255\\
\hline
0.9504 & 1.0646 & 1.0624 & 0.0357 & 0.719 & 0.7228 & 0.7191 & 0.0242\\
\hline
0.818 & 0.8414 & 0.8465 & 0.0342 & 0.6316 & 0.6599 & 0.6672 & 0.0296\\
\hline
1.1626 & 1.1685 & 1.1614 & 0.0628 & 0.6596 & 0.6332 & 0.6376 & 0.0224\\
\hline
1.0475 & 1.0098 & 1.0128 & 0.038 & 0.9136 & 0.8742 & 0.8665 & 0.0323\\
\hline
1.0174 & 0.9933 & 1.0023 & 0.0381 & 0.9257 & 0.9683 & 0.9734 & 0.0271\\
\hline
1.0666 & 1.039 & 1.0537 & 0.0463 & 0.8853 & 0.8837 & 0.893 & 0.045\\
\hline

\end{array} 
\]
\caption{Results for the maximum a posteriori learning problem for the rumours spreading model. We report mean, median, and standard deviation on 20 runs, for 10 different true parameter combinations. }
\label{table:rumoursMAP}
\end{table}

\subsubsection{Results} 
Results for maximum likelihood are shown in Table~\ref{table:rumoursML}, where we compare the true value of parameters, against the predicted mean, median, and standard deviation of the prediction in a batch of 20 runs. As we can see, the algorithm is able to reconstruct the true parameterisation with a good accuracy. As a metric to assess the quality, we consider the average observed error (euclidean distance from the true configuration), which is 0.131 for the data shown in Table \ref{table:rumoursML},  the average normalised error, obtained by dividing the absolute error by the diameter of the search space, which is 1.03\%, and the mean relative error, i.e. the absolute error for each parameter divided by the true parameter and averaged over all parameters and runs, which equals 9.23\%.
In Table \ref{table:rumoursMAP}, instead, we similarly  report the results for the maximum a posteriori estimate. In this case the average observed error is 0.074, the average normalised error is 0.58\%, and the average relative error is 5.12\%. These results  show that the use of (good) prior information can improve the performances of the algorithm, as expected, as its effect is to increase the likelihood near the optimal point and decrease it in  other areas of the parameter space.

We also run some tests to check if and to what extent the enhancements of Section \ref{sec:enhancements} are improving the search algorithm. To this end, we fixed the true parameter value to $k_s = 1.0$ and $k_r = 0.8$, we sampled 40 observations, and run the optimisation routine for  100 times, for each possible combination of the following three features: heteroschedastic noise estimation (with bootstrapping), hyperparameter optimisation, and orthogonal LHS. We then compared the distribution of the predicted parameters for each pair of combination of features, by running a Kolmogorov-Smirnoff 2 sample test, at 95\% confidence level. The p-values of the tests are reported in Table \ref{table:Pvals}. Data  show that   hyperparameter optimisation consistently and significantly improves the quality of the results.  Heteroschedastic noise estimation has a milder effect (it is  significant in 2 cases  out of four), but it relieves the user from guessing the intensity of noise.  Orthogonal sampling, instead, produces a significant improvement only in one case out of four. We can check the quality of results from the standard deviations of the predictions, shown in Table \ref{table:stdComp}: the smallest standard deviation is obtained when both heteroschedastic noise and hyperparameter optimisation were turned on.

\begin{figure}[htbp]
\begin{center}
\includegraphics[width=.47\textwidth]{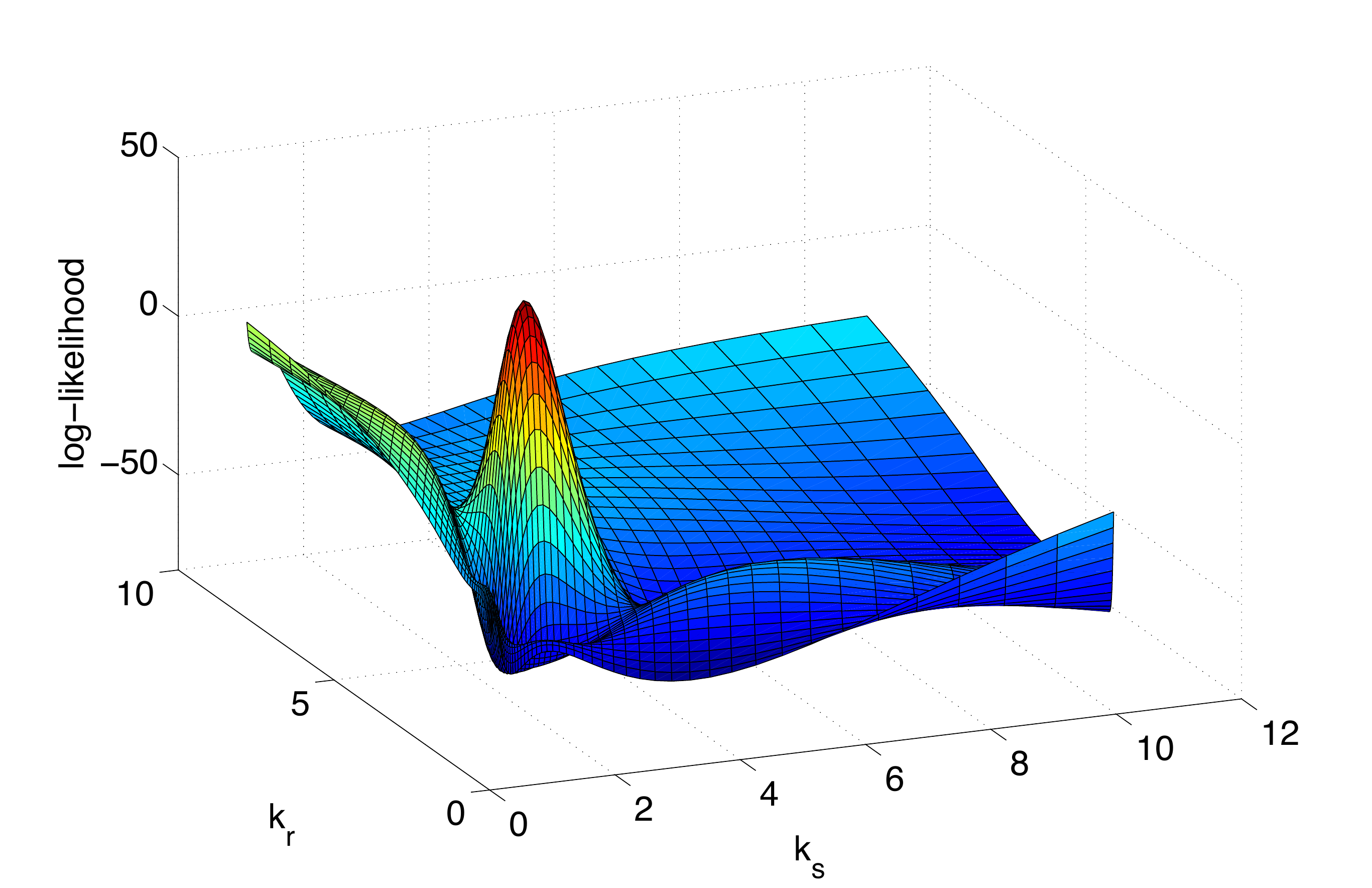}
\includegraphics[width=.47\textwidth]{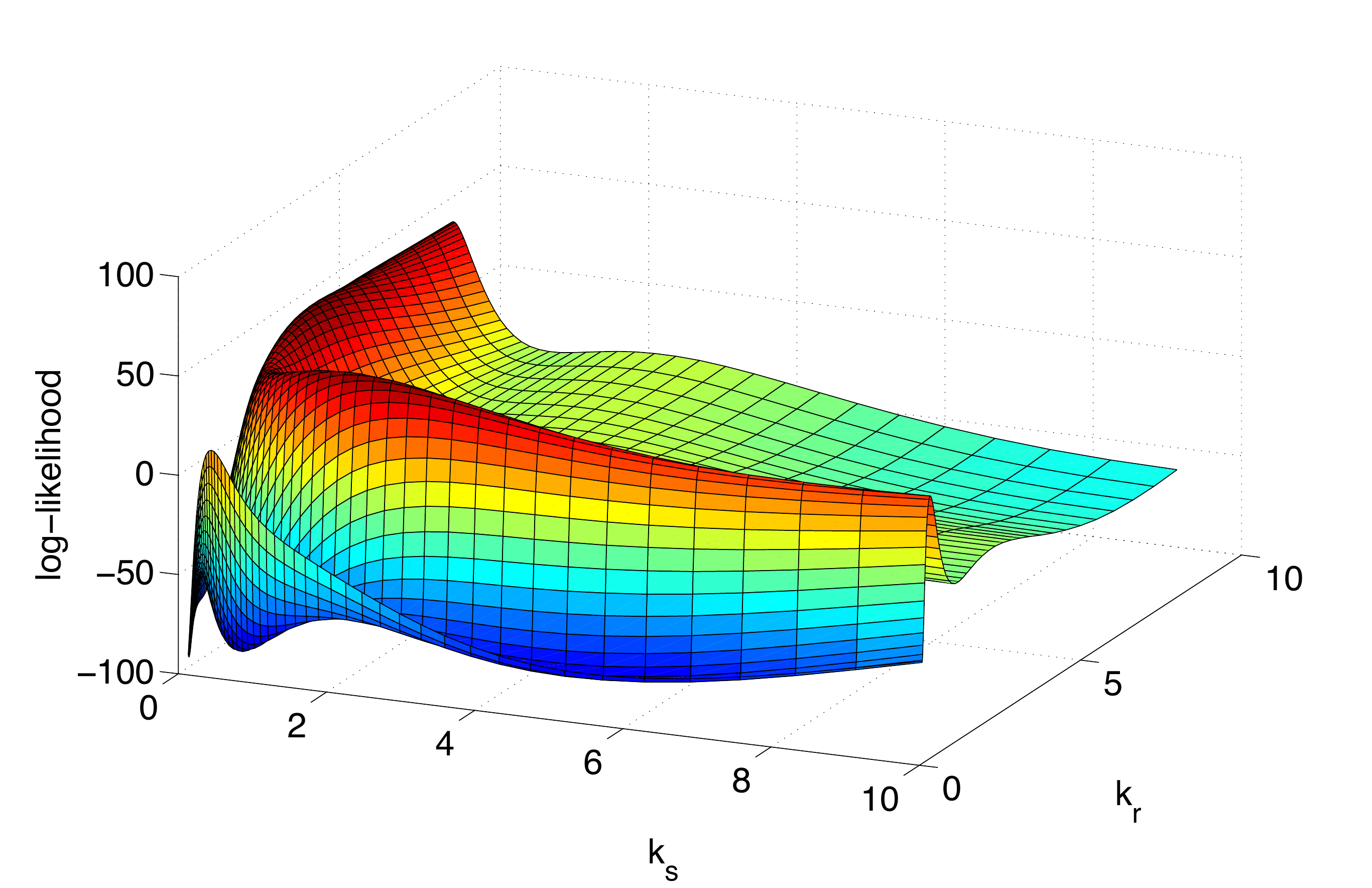}
\caption{Comparison of estimated log-likelihood surfaces for two scenarios with  true parameters fixed to $k_s = 1.0$, $k_r = 0.8$. Left:  we observe all four formulae. Right: we observe only two logical properties, the one related to extinction and the one on the stability of repressors. }
\label{fig:likelihood}
\end{center}
\end{figure}

As a final test, we consider the effect of changing the set of observable formulae, restricting to two formulae only: the one constraining the extinction time of the gossiping process, and the one concerned with the final number of people knowing the rumour. The effect of this removal is dramatic, as can be seen from Figure \ref{fig:likelihood}, where we compare the emulated log-likelihood for the full case with the emulated log-likelihood for the two formulae case. While in the first case we have a clear peak standing out, in the second setting we have an U-shaped ridge of points of almost equivalent likelihood. Not surprisingly, in this case the algorithm can  return one point from the ridge without much preference, increasing the variability of the outcome. This suggests that the choice of the logical observables is a crucial step of the method, and they should somehow capture and constrain the key features of the dynamics of the process. Further investigation on this relationship between logic and identifiability, in the light of identifying a minimal set of properties that can describe a model, is a promising future research direction.

%
%\begin{figure}[htbp]
%\begin{center}
%\includegraphics[width=.47\textwidth]{figures/noprior_pred.pdf}
%\includegraphics[width=.47\textwidth]{figures/prior_pred.pdf}
%\caption{Comparison of predicted versus true values of $k_s$ and $k_r$, for 10 different parameter values and 20 runs per parameter value. Left: ML criterion. Right: MAP criterion.}
%\label{fig:noprior_prior_pred}
%\end{center}
%\end{figure}
%
%
%\begin{figure}[htbp]
%\begin{center}
%\includegraphics[width=.5\textwidth]{figures/predictions.pdf}
%\caption{Predicted values (blue crosses) and true parameter (red x) configuration in the $k_s-k_r$-plane, for 100 runs of the optimisation algorithm.}
%\label{fig:pred}
%\end{center}
%\end{figure}

\begin{table}[!t]
\[
\begin{array}{r|cccccccc}
 & 001 & 010 & 011 & 100 & 101 & 110 & 111\\
 \hline
000 & \mathbf{< 10^{-5}}    & 0.3439    &\mathbf{< 10^{-5}}    & \mathbf{0.0314}    &\mathbf{< 10^{-5}}    & 0.1400    &\mathbf{< 10^{-5}}\\
001     &    & \mathbf{< 10^{-5}}    & \mathbf{0.0082}    &\mathbf{< 10^{-5}}    & 0.3439    &\mathbf{< 10^{-5}}    & 0.6766\\
010 &            &    &\mathbf{< 10^{-5}}    & 0.2606    &\mathbf{< 10^{-5}}    & 0.8938    &\mathbf{< 10^{-5}}\\
 011 &           &        &    &\mathbf{< 10^{-5}}    & 0.1930    &\mathbf{< 10^{-5}}    & \mathbf{0.0314}\\
 100  &         &        &        &    &\mathbf{< 10^{-5}}    & 0.4431    &\mathbf{< 10^{-5}}\\
 101     &      &        &        &        &    &\mathbf{< 10^{-5}}    & 0.4431\\
 110   &        &        &        &        &        &    &\mathbf{< 10^{-5}}\\
\end{array}
\]
\caption{P-values for the two sample Kolmogorov-Smirnoff test for the comparison of the different enhancements discussed in Section \ref{sec:enhancements}. We compared the predicted values of parameter $k_s$. Results for $k_r$ are similar.  The three-digit labels of rows and columns refer to the presence (1) or absence (0) of a specific feature in the optimisation. The first digit from the left is the estimation of heteroschedastic noise, the second is the orthogonal grid sampling, the third is the hyperparameter optimisation. Significant values at 95\% confidence are in bold.  }
\label{table:Pvals}
\end{table}

\begin{table}
\[
\begin{array}{|c|c|c|c|c|c|c|c|c|}
\hline
              &   000    &       001   &    010     &    011         &    100    &    101       &  110      &    111      \\
              \hline    
std(k_s) & 0.0319 &    0.0203 &    0.0333 &    0.0197   & 0.0315 &    0.0200   & 0.0312 &    0.0196 \\
\hline
std(k_r) & 0.0338 &   0.0203   & 0.0352 &    0.0196 &    0.0320 &    0.0202   & 0.0326 &    0.0196\\
\hline
\end{array}\]
\caption{Standard deviations of predicted parameter values  for the comparison of the different enhancements discussed in Section \ref{sec:enhancements}. The three-digit labels of  columns are as in the caption of Table \ref{table:Pvals}.  }
\label{table:stdComp}
\end{table}

\subsection{Genetic Toggle-Switch}
\label{sec:toggle}
We consider now a model of a simple genetic network implementing a toggle-switch, i.e. a form of local memory \cite{gardner2000construction,Tian2006}. The gene circuit is composed of two genes $G_1$ and $G_2$, expressing proteins $X_1$ and $X_2$ that act as mutual repressors: $X_1$ represses $G_2$ and $X_2$ represses $G_1$. For certain values of the parameter space, this circuit has two stable states, one in which the first protein is expressed and the second is not, and the symmetric one. Internal noisy fluctuations or external stimuli, like an increase in temperature, can force the system to jump from one stable state to the other. Hence, a proper treatment of stochastic behaviour is fundamental to properly understand (and design) this kind of circuit. 
The model we consider here follows the approach of \cite{Ocone2013}, and describes the genetic network as a stochastic hybrid system, where genes are modelled as a two-state telegraph processes, while proteins are represented as continuous species, subject to a noisy continuous evolution given by an SDE with drift modulated by the state of the gene. More specifically, protein $X_i$ evolves according to the SDE
\[ dX_i = (\lambda_i G_i - \mu_ix_i)dt + \sigma_i dW_i,\]
while gene $G_i$ changes from state $G_i=1$ to $G_i = 0$ with rate $f_i^- = k_i\exp(\alpha_i X_j )$, $j\neq i$, and jumps back to the active state with constant rate $c_i$.
The toggle-switch genetic network is known to be bistable. The two stable equilibria correspond to one protein expressed and the other not expressed. Hence, we consider four MiTL formulae, two per protein, expressing the active and inactive status. Furthermore, we require the protein to remain active or inactive for some time. Specifically, the formulae we consider are
\begin{enumerate}
\item $\eventually{0}{T}\always{0}{T_1}\eventually{0}{T_2} X_i\geq th_{high}$, expressing the fact that between time $[0,T]$ the system stabilises for $T_1$ time units in a state in which $\eventually{0}{T_2} X_i\geq th_{high}$ holds, i.e. a state such that  protein $X_i$ is always found above $th_{high}$ within additional $T_2$ time units.\footnote{Notice that we do not require the protein to remain constantly above $th_{high}$, because we will choose a large value for the threshold, and the noisy evolution will easily make the protein fall below $th_{high}$ in $T_1$ time units.}
\item $\eventually{0}{T}\always{0}{T_1} X_i\leq th_{low}$, expressing the fact that $X_i$ remains inactive (below $th_{low}$) for $T_1$ time units. 
\end{enumerate}
If all formulae are true for both proteins, we are in a situation in which the process jumps from one  stable state to the other during its observed life span of $T$ time units.  

\subsubsection{Experimental Setup}
We consider a scenario in which genes are symmetric, having a total of six parameters: two for the protein dynamics ($\lambda$ and $\mu$), three for the telegraph process ($k$, $c$, and $\alpha$), and one for the noise ($\sigma$). In the experiments, we decided to fix the production rate $\lambda = 2$ and the degradation rate $\mu = 0.01$, exploring the other four parameters. As for the rumour spreading model, we consider 40 observations, generated from random parameters values sampled uniformly according to $k\in[0.08,0.12]$, $c\in[0.03,0.07]$, $\alpha\in[0.08,0.12]$, and $\sigma\in[0.8,1.2]$. For each of the 5 parameter combinations generated, we run 6 times the optimisation of the log-likelihood. Parameters were searched in the space $k\in[0.01,1.0]$, $c\in[0.005,0.5]$, $\alpha\in[0.01,1.0]$, and $\sigma\in[0.1,10]$, after log-standardising parameter ranges. Parameters of the formula where set as follows: $th_{low} = 20$, $th_{high} = 80$ (the average concentration of a protein in absence of regulation is 200), $T=7000$, $T_1 = 1000$, and $T_2 = 200$. In particular, notice that we look at a very long temporal window, and require properties to hold for a long time.
We run the GP-UCB optimisation starting from 96 points sampled using the orthogonal LHS scheme, and we evaluated the emulated function on a random grid of 1024 points, again sampled according to LHS. We used bootstrapping estimation of (heteroschedastic) noise, and we optimised hyperparameters at each run.

\subsubsection{Results}
\begin{figure}[htbp]
\begin{center}
\includegraphics[width=.32\textwidth]{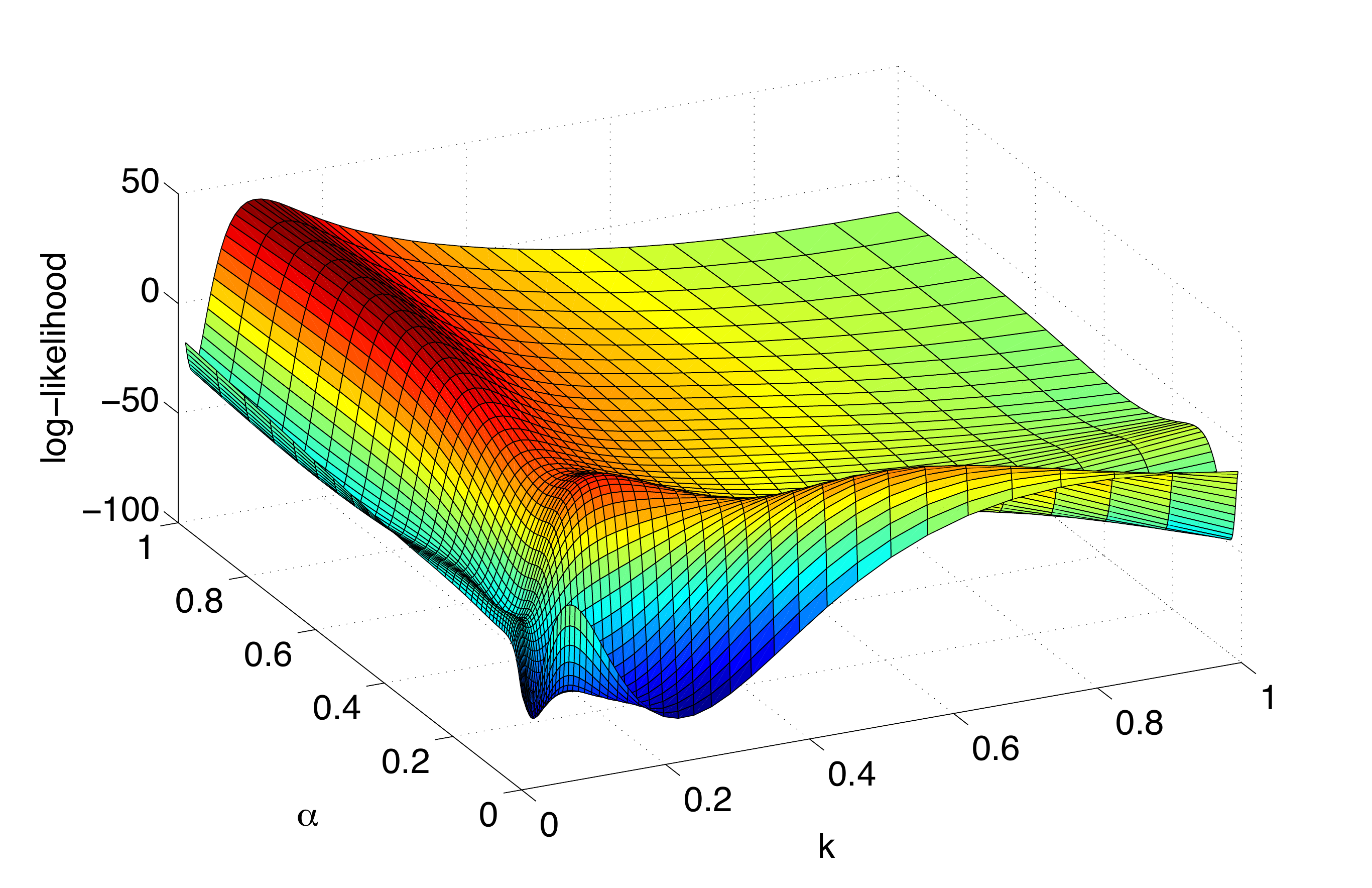}
\includegraphics[width=.32\textwidth]{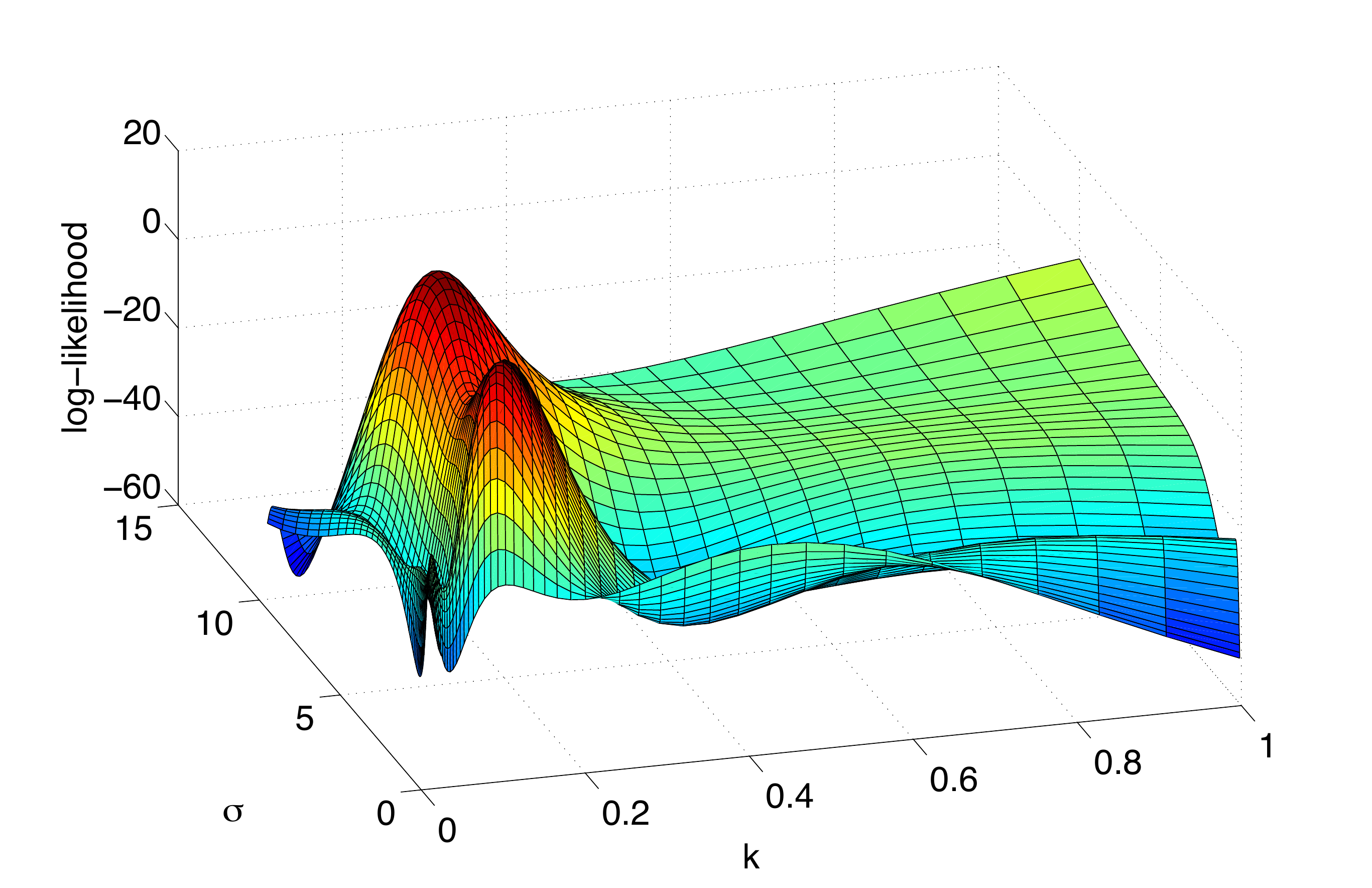}
\includegraphics[width=.32\textwidth]{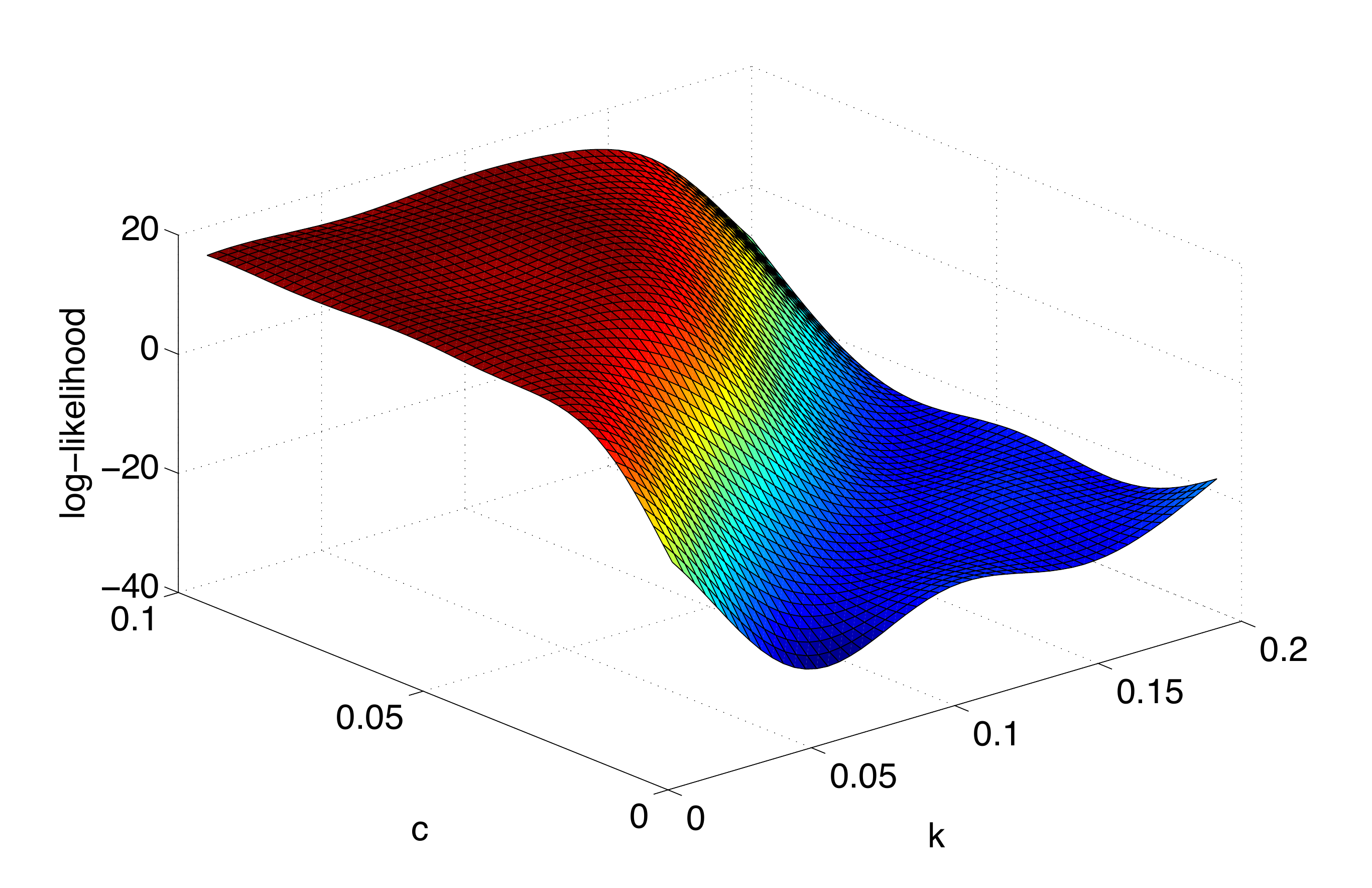}
\caption{Emulation of log-likelihood as a function of $k$ and $\alpha$ (left), $k$ and $\sigma$ (middle), and $k$ and $c$ (right) in the toggle switch example.}
\label{fig:toggleBAD}
\end{center}
\end{figure}

The results for the exploration of 4 parameters are reported in Table \ref{table:toggleML}. We obtained an average distance from the true parameter configuration of 0.8537 and an average normalised distance of 0.0853. Inspecting the data in more detail, we can see that parameters are captured less  accurately than in the rumour spreading case and that there is a remarkable variability between the simulation runs. To better validate the accuracy of the data, in Table  \ref{table:toggleML} we also show the mean value of the standard deviation estimated by the Laplace method. If we consider the fraction of optimisation runs in which the true parameter value falls within the 95\% confidence interval constructed using the Laplace approximation (data not shown, but derivable from Table  \ref{table:toggleML}), we can observe that the parameters captured more accurately are $k$ and $c$, as they almost always fall within the 95\% confidence interval. The estimate of $\alpha$ and $\sigma$, instead, are subject to a much larger variability. This is a sign of a rugged log-likelihood landscape. To understand the origin of such a behaviour and check if it is caused by an intrinsic lack of identifiability of some of the parameters, given the observed data, we rerun the optimisation on different subsets of two parameters, fixing the other two to a nominal value of $k = 0.1$, $c=0.05$, $\alpha = 0.1$, and $\sigma=1$. What we observed is reported in Figure \ref{fig:toggleBAD}, in which the left and middle charts show the estimated log-likelihood surface for $k$ and $\alpha$ and for $k$ and $\sigma$. In both cases, we see a ridge or multiple maxima aligned on a line parallel to the $\alpha$ or $\sigma$ axes, of approximatively constant height. This basically shows that the system is largely insensitive to the precise value of $\alpha$ and $\sigma$, provided they remain within a reasonable range. On the other hand, $k$ is predicted reasonably accurately. This is in agreement with \cite{Ocone2013}, where insensitivity with respect to $\alpha$ has also been observed. In the right chart of Figure \ref{fig:toggleBAD}, instead, we show the estimated log-likelihood as a function of $k$ and $c$, varying them in the region $k\in[0.01,0.2]$, $c\in[0.005,0.1]$. As we can see, there is a flat region in the upper left corner, corresponding to small values of $k$ and $c$. This shows that $k$ and $c$, for the current formulae, cannot be identified precisely, explaining the results in Table~\ref{table:toggleML}. Note that this region is nonetheless relatively small, and the Laplace approximation manages to capture, at least partially, the variability in the prediction. 
We note here that looking at the 2 dimensional landscape of the log-likelihood, for pairs of parameters, can be a potentially interesting direction to investigate, in order  to get insights on parameter identifiability, and to infer possible relationships between parameters, also to reduce the search space.  This can be also combined with sensitivity analysis, to identify the most relevant parameters to explore.

\begin{table}
\[{\small
\begin{array}{|c|c|c|c||c|c|c|c|}
\hline
\text{true $k$}  & \text{mean $k$ $\pm$ std} & \text{median $k$} & \text{Lap. sdt $k$} & \text{true $\alpha$} & \text{mean $\alpha$ $\pm$ std } & \text{median $\alpha$} & \text{Lap. std $\alpha$}\\
\hline
 0.09   & 0.035 \pm  0.017   & 0.034   &  0.054  & 0.112   & 0.201 \pm 0.167   & 0.12   & 0.058 \\
 \hline
    0.103   & 0.034 \pm 0.023  & 0.028  &  0.058  & 0.094   & 0.309 \pm 0.233  & 0.247   & 0.067 \\
    \hline
    0.099   & 0.067 \pm 0.05  & 0.053   & 0.052  & 0.12   & 0.384  \pm 0.321 & 0.257   & 0.071\\
    \hline
    0.107   & 0.045 \pm0.044  & 0.023   & 0.054  & 0.098   & 0.218 \pm 0.262   & 0.082   & 0.044\\
    \hline
    0.085   & 0.018 \pm 0.004   & 0.017   &  0.053 & 0.091   & 0.336 \pm 0.176   & 0.308   & 0.066 \\
\hline
\hline
\text{true $c$}  & \text{mean $c$ $\pm$ std} & \text{median $c$} & \text{Lap.  std $c$} & \text{true $\sigma$} & \text{mean $\sigma$ $\pm$ std} & \text{median $\sigma$} & \text{Lap std $\sigma$}\\
\hline
    0.062   & 0.073 \pm 0.034  & 0.066  &  0.054  & 1.2   & 1.168 \pm 1.123  & 0.978   & 0.06 \\
\hline
    0.062   & 0.09 \pm 0.107  & 0.056    & 0.074 & 0.834   & 1.512 \pm 1.048   & 1.79   & 0.074 \\
\hline
    0.041   & 0.018 \pm 0.015   & 0.012 & 0.056 & 0.837   & 1.134 \pm 1.042   & 0.877   & 0.078 \\
\hline
    0.07     & 0.1 \pm 0.048  & 0.088     & 0.066 & 1.161   & 0.817 \pm 0.446   & 0.862   & 0.061 \\
\hline
    0.058   & 0.025 \pm 0.01  & 0.022   & 0.056 & 1.044   & 0.795  \pm 0.765 & 0.525   & 0.072 \\
\hline
\end{array} 
}\]
\caption{Results for the maximum likelihood learning problem for the toggle switch model, for the joint optimisation of $k$, $c$, $\alpha$ and $\sigma$. We report mean plus/minus standard deviation, median, and mean standard deviation estimated by the Laplace method.  We consider 6 runs, for 5 different true parameter combinations. }
\label{table:toggleML}
\end{table}

\subsubsection{System Design}
To test the performances of the method for the design problem, we consider again the toggle switch scenario, and the formulae described in the previous section, only for one protein. Notice that one formula describes a state in which the protein is expressed, while the other a state in which the protein is repressed. They can be both true in the same trajectory only if  the system jumps from one stable state to the other within time $T=7000$. In this experiment, we will try to force this phenomenon not to happen, at least for within time $[0,T]$. This can be obtained by a distribution  of truth values putting mass 0.5 on the situation in which the first formula is true and the second is false, and 0.5 on the symmetric case. The choice of this distribution is due to the symmetry of the system, and the fact that we start simulations from a symmetric state, hence we expect a symmetry in the probability distribution of the truth of the two formulae.
With this  target probability in mind, we run the optimisation of the JSD exploring a space of  three parameters: the production rate $\lambda\in[0.2,20]$, the binding strength $k\in[0.01,1]$, and the unbinding rate $c\in[0.01,1]$. We run the optimisation 25 times, obtaining an average JSD of 0.0011. The probability distribution obtained match the targeted probabilities quite well:  the mean difference between probability values is of 0.011, while the max difference is 0.026. The mean, median and the standard deviation of the parameters returned by the optimisation are shown in Table \ref{table:toggleDesign}, where we can see that there is a large variability on $\lambda$, meaning that the parameter is not very important for the design task, provided it is large enough, and a much small variability in $k$ and $c$. In particular, $k$ here is consistently smaller than $c$, and this seems a crucial aspect in matching the design specification.

\begin{table}
\[
\begin{array}{|c|c|c|c|c|}
\hline
  parameter & mean &  median & std & Laplace\ std\\
    \hline
\lambda &    2.9651   & 1.9561   &  2.0725  &  0.2432\\
    \hline
  k &   0.0946   & 0.0736   &  0.0704  &  0.2856\\
    \hline
   c & 0.4307   & 0.3606  &   0.2364  &  0.3096\\
    \hline
\end{array} 
\]
\caption{Results for the design problem for the toggle switch model, for the joint optimisation of $k$, $c$, and $\lambda$. We report mean, median, and standard deviation of 25 optimisation runs plus the mean standard deviation estimated by Laplace method. }
\label{table:toggleDesign}
\end{table}

\section{Conclusions}
\label{sec:conc}

The role of uncertainty in formal modelling has historically been a controversial one: while stochastic processes are now common modelling tools in computer science \cite{Baier2008}, much less work has been dedicated to stochastic models with parametric uncertainty. In this paper we argue that considering parametric classes of stochastic models can be a natural scenario in many real applications, and explore how advanced verification tools can be coupled with ideas from machine learning to yield effective tools for integrating logical constraints in modelling and design tasks. 

Our paper is part of a growing family of works which attempt to bring tools from continuous mathematics and machine learning into formal modelling. The main reference for this paper is the earlier conference paper \cite{Bortolussi2013}: this is considerably expanded in this work, in particular by providing an automated procedure to set all the parameters involved in the framework. Related ideas which embed a stochastic model in a local family have been explored in the context of analysing the robustness of logical properties \cite{Bartocci2013} and in the context of model repair \cite{Bartocci2011}. \guido{GP optimisation in a formal modelling context has also been employed in the converse problem of learning temporal logic specifications from data \cite{Bartocci:data14}, and is the basis of a recent novel approach to reachability computations \cite{Bortolussi:statistical14}.}

The methodology presented in this paper offers both a promising avenue to tackle practically relevant modelling problems, and intriguing further challenges. \guido{A natural extension of our work would be to consider the identification of model structures, as opposed to model parameters only. While in principle straightforward, algorithmic adjustments will be needed to enforce sparsity constraints in the optimisation.} From the modelling point of view, the idea of performing system identification from logical constraints immediately begs the question of a minimal set of logical properties that enable the identification of a system within a parametric family. We don't have an answer to this question, but it is likely that ideas from continuous mathematics will be useful in further exploring this fascinating question. From the computational point of view, the methods we use are limited to exploring systems with a handful of parameters. Scaling of Bayesian optimisation algorithms is a current topic of research in machine learning, and innovative novel ideas on randomisation \cite{Wang:Bayesian13} may hold the key to applying these methodologies to large scale formal models.

\bibliographystyle{plain}
\bibliography{pcbib}

\end{document}